\documentclass[preprint,superscriptaddress,amsmath,amssymb,aps,pre,floatfix,a4paper]{revtex4-2}

\usepackage{graphicx}
\usepackage{latexsym}
\usepackage{bm}
\usepackage{amsmath}
\usepackage{amsfonts}
\usepackage{amssymb}
\usepackage{enumitem}

\usepackage[usenames,dvipsnames]{xcolor}

\renewcommand{\vec}[1]{\underline{#1}}                      
\newcommand{\Tg}{\ensuremath{T_\text{g}}}                   

\begin{document}

\title{The constrained molecular dynamics (CMD) to obtain spatio-temporal memory functions of viscoelastic liquids: stability analysis of equations of motion}

\author{O. Benzerara}
\affiliation{Universit\'e de Strasbourg, CNRS, Institut Charles Sadron UPR 22, F-67000 Strasbourg, France}

\author{J. Baschnagel}
\affiliation{Universit\'e de Strasbourg, CNRS, Institut Charles Sadron UPR 22, F-67000 Strasbourg, France}

\author{A.N. Semenov *}
\affiliation{Universit\'e de Strasbourg, CNRS, Institut Charles Sadron UPR 22, F-67000 Strasbourg, France}



\date{October 27, 2026}

\begin{abstract}
Mechanical stress is one of the main characteristics of viscoelastic molecular systems,
including vitrifying liquids and amorphous solids which are known to show long-range dynamical heterogeneities. 
It is therefore important to deal with time- and space-resolved local stress whose correlations can be used to obtain viscoelastic memory functions of such systems, in particular, their wave-vector dependent shear and compression relaxation moduli. However, such memory functions are not directly accessible within standard molecular dynamics (MD) 
simulations.
Here we discuss a new method called the constrained molecular dynamics (CMD) which allows for a direct evaluation of
the wave-vector and time 
dependent viscoelastic moduli. The study is focused, in particular, on the stability analysis of different variants of the CMD, revealing the most efficient stabilization algorithm. Our theoretical predictions for the time-dependence of the most important parameters (like the wave-vector resolved mass current) during the stabilization stage are compared with the simulation data using different CMD algorithms. An excellent agreement is observed, thus confirming the validity of our theoretical approach and its ability to accurately predict the most important initial part of CMD trajectories.        
\end{abstract}


\maketitle


\section{Introduction}
Mechanical stress is a fundamental rheological quantity in supercooled liquids, complex fluids, and amorphous solids \cite{BerthierBiroliRMP,Fuchs:AdvPolymSci2010,NicolasEtal:RMP2018}. In such systems, the flow and deformation at a given time generally depend on the prior stress history. In nonequilibrium amorphous solids, residual stresses generated during preparation have an impact on mechanical properties, including fragility, robustness, and impact resistance \cite{Voigtmann:COCIS2014}. Moreover, time-dependent correlations of the volume-averaged stress can be used to obtain key viscoelastic memory functions of supercooled and complex liquids \cite{AllenTildesley,HansenMcDonald,GoetzeBook2009}, including the shear and bulk relaxation moduli \cite{KriuchevskyiEtal:PRL2017,KlochkoEtal:SoftMatter2021,xx2}.

In the past decade, many simulation studies have examined spatially resolved stress correlations in amorphous systems  \cite{Lemaitre:PRL2014,Lemaitre:JCP2015,Lemaitre:PRE2017,Lemaitre:JCP2018,WuEtal:PRE2015,ChowdhuryEtal:JCP2016,TongSenguptaTanaka:NatComm2020,FurukawaTanaka:PRL2009,MaierEtal:JCP2018,VogelFuchs:EPJE2020,SteffenEtal:JCP2022,KlochkoEtal:JCP2022,WittmerEtal_P5a:PRE2023,Grimm2024}. Remarkably, even in isotropic liquids with short-ranged structural pair correlations, the correlations of the shear stress exhibit angle-dependent, long-range spatial decay, analogous to that found in amorphous solids \cite{Lemaitre:PRE2017,Lemaitre:JCP2018}. As emphasized by Lemaître \cite{Lemaitre:JCP2015}, this angular dependence does not imply an underlying structural anisotropy. Rather, it follows from the tensorial nature of stress~\cite{Lemaitre:JCP2015,Lemaitre:PRE2017,MaierEtal:JCP2018,KlochkoEtal:SoftMatter2018,WittmerEtal_P5a:PRE2023,SemenovBaschnagel:Polymers2024} in spite of macroscopic symmetries of the amorphous systems (invariance relative to space and time translations, isotropy, achirality). Such long-range stress correlations are therefore expected not only in glasses but, more generally, in viscoelastic materials, including dynamic polymer networks \cite{MartinsEtal:Macro2023} and polymer melts \cite{Likhtman:Review2012,SemenovEtal:JCP2012}. Their emergence raises fundamental questions regarding their role in structural relaxation or dynamic heterogeneities in glassy systems whose range increases 
on cooling toward the glass transition temperature $T_g$ and below
 \cite{ChackoEtal:PRL2021,OzawaBiroli:PRL2023,164505.17Deben,N16,N19,N21,N6,N24,N5}.

One way to elucidate this issue is to examine not only the spatial dependence but the spatio-temporal dependence of the correlations of the full stress tensor. From a theoretical perspective, this problem has been approached within two distinct frameworks: the Zwanzig--Mori (ZM) projection-operator formalism \cite{MaierEtal:PRL2017,MaierEtal:JCP2018,MaierEtal:JCP2024,VogelFuchs:EPJE2020} and an approach based on the fluctuation-dissipation theorem (FDT) and the new concept of ``stress noise'' \cite{KlochkoEtal:SoftMatter2018,SemenovBaschnagel:Polymers2024,BaschnagelSemenov:SoftMatter2026,JCP2022}. Within the ZM formalism, the time-dependent equilibrium correlation function $C$ of the stress tensor $\sigma_{\alpha\beta}(\underline q,t)$ at wavevector $\underline q$ and time $t$ ($\alpha,\beta$ denote Cartesian components) is related to a memory kernel $M$ that encodes all non-Markovian effects governing the relaxation of $\sigma_{\alpha\beta}(\underline q,t)$~\cite{MaierEtal:JCP2018,MaierEtal:JCP2024,VogelFuchs:EPJE2020}. Although the relation between $C$ and $M$ is exact, further analysis is complicated by the fact that the time evolution of $M$ is generated by an unusual non-Newtonian dynamics intrinsic to the ZM formalism \cite{GoetzeBook2009,Janssen2018,PihlajamaaEtal:preprint2023,PihlajamaaJanssen:PRR2024,Baity-JesiReichman:JCP2019}. 
 The second approach~\cite{KlochkoEtal:SoftMatter2018,SemenovBaschnagel:Polymers2024,BaschnagelSemenov:SoftMatter2026,JCP2022} is conceptionally based on the fact that $\sigma_{\alpha\beta}(\underline q,t)$ can be decomposed into two parts: one part that is linked to the wave-vector and time dependent flow field and another part that is decoupled from the flow and is termed ``stress noise''.  The theory underlying the second 
approach confirms the long-range character of stress correlations in 
glassy systems and was shown to be in excellent agreement with 
simulation data on two-dimensional systems of polydisperse Lennard--Jones (LJ) particles (2D pLJ) 
at $q\alt 0.08q_{max}$ (where $q_{max}$ corresponds to the main peak of the static structure factor)~\cite{JCP2022}.
This theory was then further generalized to yield the general relations 
(valid at any $q$) between the tensorial correlation function of the 
stress, $C_{\alpha\beta\gamma\delta}(\underline q,t)$, and the tensor of generalized relaxation moduli, 
$E_{\alpha\beta\gamma\delta}(\underline q,t)$, on the one hand, and between the $q,t$-resolved elasticity 
tensor and the correlation functions of the stress noise, on the other 
hand~\cite{Polymers2024}. Finally, it was recently shown that 
the two approaches (based on the ZM formalism and on the `stress 
noise' concept) are in complete agreement with each other~\cite{SM2025}. 
In particular, it was shown that the memory kernel of ZM formalism is identical to the time-dependent correlation function of the stress noise. This result makes it possible, in principle, to determine the spatio-temporally resolved memory kernel directly from molecular-dynamics (MD) trajectories because the stress-noise correlation function can be accessed by a recently introduced MD approach in which, at each time step, a constraint is imposed that suppresses the mass current at wave vector $\underline q$. We recently presented a brief account of this ``constraint MD'' (CMD) method, described its implementation within the velocity-Verlet algorithm, and provided numerical evidence for its validity for a two-dimensional glass-forming system \cite{BenzeraraSemenovBaschnagel:PRL2026}. In the present work, we provide a more detailed theoretical and numerical analysis of the CMD method. 

Our paper is structured as follows. In Sec.~II, we explain why constrained molecular dynamics is required. The fundamental equations underlying CMD are derived in Sec.III. On this basis, we derive several approximate iterative schemes for implementing the CMD dynamics and analyze their stability in Secs.~IV and V. The effects of subdominant internal forces and other corrections to the leading approximation are discussed in Sec.VI. Section VII concludes with a summary.

\section{The purpose of the CMD approach}

Let us consider a (generally polydisperse) system of $N$ spherical 
particles in volume $V$.  The system is supposed to be macroscopically 
uniform, isotropic and fluid-like.  The typical real examples can be 
provided by supercooled liquids or colloidal solutions.  

At each time instant the system microstate is characterized by mass 
density, $\rho (\underline r,t)$, and momentum density (mass current density), $\underline 
j(\underline r,t)$, fields:  
\begin{equation}\rho (\underline r,t)=\sum_am_a\delta (\underline r-\underline r_
a(t))\label{f1}\end{equation}
\begin{equation}\underline j(\underline r,t)=\sum_am_a\underline v_a(t)\delta (\underline 
r-\underline r_a(t))\label{f2}\end{equation}
where $m_a$, $\underline r_a(t)$ and $\underline v_a(t)$ are respectively the mass, position and 
velocity of particle `$a$' at time $t$ (and the sum runs over all 
particles).  The associated `weighted' concentration and velocity fields 
are 
\begin{equation}c(\underline r,t)=\rho (\underline r,t)/\bar {m},~~\underline v(\underline 
r,t)=\underline j(\underline r,t)/\rho_0\label{e1a}\end{equation}
where $\bar {m}=\sum_am_a/N$ is the mean particle mass, $\rho_0=N\bar {m}/V$ is the 
mean mass density. The two fields are related by the mass 
conservation law: 
\begin{equation}\frac {\partial\rho (\underline r,t)}{\partial t}=-\nabla\cdot\underline 
j(\underline r,t)\label{f4}\end{equation}
In turn, the rate of change of the mass current density $\underline j$ is defined by the 
fundamental momentum equation involving the local 
stress tensor field $\sigma_{\alpha\beta}(\underline r,t)$~\cite{TadMil,BalZop}:
\begin{equation}\frac {\partial j_{\alpha}(\underline r,t)}{\partial t}=\frac {\partial}{
\partial r_{\beta}}\sigma_{\alpha\beta}(\underline r,t)\label{f5a}\end{equation}
where $\alpha$, $\beta$ indicate the Cartesian components (and summation over 
repeated subscripts is assumed).  It is convenient to deal with 
Fourier transforms of the relevant fields defined like 
\begin{equation}\sigma_{\alpha\beta}(\underline q,t)=\mbox{\rm FT}[\sigma_{\alpha
\beta}(\underline r,t)]\equiv\frac 1V\int\sigma_{\alpha\beta}(\underline r,t)e^{
-{\rm i}\underline q\cdot\underline r}{\rm d}^dr\label{Fsig}\end{equation}
where $d$ is the space dimension. 
The microscopic equation defining the stress tensor $\sigma_{\alpha\beta}(\underline 
q,t)$ in terms 
of particle positions and interactions is given in eq.  13 of 
ref.~\cite{SM2025} (see also refs.~\cite{TadMil,BalZop}):
  \begin{equation}
  	\sigma_{\alpha\beta}(\underline q,t)
  	=  \frac 1V\sum_{a>b}
  	\frac{F_{ab\alpha}^\text{int}r_{ab\beta}}{{\rm i}\underline 
  		q\cdot\underline r_{ab}} \left ( {\rm e}^{-{\rm i}\underline 
  		q\cdot\underline r_a}-{\rm e}^{-{\rm i}\underline q\cdot\underline r_b}\right )
  	-\frac 1V\sum_a m_a v_{a\alpha}v_{a\beta}{\rm e}^{-{\rm i}\underline 
  		q\cdot\underline r_a} , 
  	\label{G2}
  \end{equation}
  where the first sum includes all pairs $a,b$ of interacting particles, $\underline r_{ab} = \underline r_b - \underline r_a$, and $F_{ab\alpha}^\text{int}$ is the component $\alpha$ of the interaction force, $F_{ab\alpha}^\text{int} \equiv (\frac {r_{\alpha}}ru'_{ab}(r))_{\underline r=\underline r_b-\underline r_a}$, with $u_{ab}(r)$ being 
  the pairwise interaction energy of particles $a$ and $b$, and $u'_{ab}(r)\equiv\frac {{\rm d} u_{ab}(r)}{{\rm d}r}$.
 
Replacing $\sigma_{\alpha\beta}(\underline r,t)$ in the equation~\ref{Fsig} with $\rho 
(\underline r,t)$ or $\underline v(\underline r,t)$ 
and using 
eqs.~\ref{f1},~\ref{f2},~\ref{e1a} we get the collective mass density 
$\rho (\underline q,t)$ 
and the collective velocity $\underline J(\underline q,t)\equiv\underline v(\underline 
q,t)$ fields in 
Fourier space:  
\begin{equation}\rho (\underline q,t)=\mbox{\rm FT}[\rho (\underline r,t)]=\frac 
1V\sum_am_ae^{-{\rm i}\underline q\cdot\underline r_a(t)}\label{f8}\end{equation}
\begin{equation}\underline J(\underline q,t)=\mbox{\rm FT}[\underline v(\underline 
r,t)]=\frac 1{N\bar {m}}\sum_am_a\underline v_a(t)e^{-{\rm i}\underline q\cdot\underline 
r_a(t)}\label{e1}\end{equation}
(Note that here and below the collective velocity is denoted as $\underline J(\underline 
q,t)$ 
instead of $\underline v(\underline q,t)$ for its better visibility in equations below.)

The basic conservation laws in terms of the transformed fields read
\begin{equation}\frac 1{\rho_0}\frac {\rho (\underline q,t)}{\partial t}=-{\rm i}
q_{\alpha}J_{\alpha}(\underline q,t)\label{f6}\end{equation}
\begin{equation}\rho_0\frac {\partial J_{\alpha}(\underline q,t)}{\partial t}={\rm i}
q_{\beta}\sigma_{\alpha\beta}(\underline q,t)\label{f7}\end{equation}

The tensor of generalized relaxation moduli (referred to also as the `elasticity tensor'), $E_{\alpha\beta\gamma\delta}(\underline 
q,t)$, defines the ensemble-averaged
stress response to a small deformation tensor $\dot{\gamma}_{\alpha'\beta'}(\underline 
q,t'){\rm d}t'$ at a prior 
time $t'<t$, and eventually the total increment of the stress,
$\Delta\sigma_{\alpha\beta}(\underline q,t)$, due to a weak flow in the system:~\cite{Polymers2024,SM2018,JCP2022} 
\begin{equation}\left<\Delta\sigma_{\alpha\beta}(\underline q,t)\right>=\int_{-\infty}^{\infty}
E_{\alpha\beta\alpha'\beta'}(\underline q,t-t')\left<\dot{\gamma}_{\alpha'\beta'}(\underline 
q,t')\right>{\rm d}t'\label{ee27}\end{equation}
where $E_{\alpha\beta\alpha'\beta'}(\underline q,t)=0$ at $t<0$ due to causality, and an equilibrated isochoric 
ensemble is assumed.  Here $\dot{\gamma}_{\alpha\beta}$ is the 
tensorial deformation rate whose real-space components are velocity gradients,  
$\dot{\gamma}_{\alpha\beta}(\underline r,t)=\partial v_{\alpha}(\underline r,t)/
\partial r_{\beta}$. Hence its Fourier transform is (cf eq.~\ref{e1}):  
\begin{equation}\dot{\gamma}_{\alpha\beta}(\underline q,t)={\rm i}q_{\beta}J_{\alpha}
(\underline q,t)\label{gvr}\end{equation}
Note that (i) the velocity field $\underline v(\underline r,t)$ in real space and its Fourier 
transform $J(\underline q,t)$ have the same physical dimension, (ii) 
$\underline J(-q,t)=\underline J^{*}(q,t)$ since $\underline q$, $\underline r_a$ and $\underline 
v_a$ are always real, and (iii) the flow field $v_{\alpha}(\underline r,t)$  can 
be generated by an external force field as discussed in 
ref.~\cite{Polymers2024} and below.  

The tensorial correlation function of stress is defined 
as (for $q\neq 0$)~\cite{SM2018,Polymers2024}: 
\begin{equation}C_{\alpha\beta\alpha'\beta'}(\underline q,t)\equiv\frac VT\left<
\sigma_{\alpha\beta}(\underline q,t+t')\sigma_{\alpha'\beta'}(-\underline q,t')\right
>\label{e300}\end{equation}
where $T$ is temperature in energy units (hence Boltzmann constant $k_\text{B}$ is omitted).  The angular brackets $\left<...\right
>$ 
indicate both averaging over an equilibrated statistical ensemble (eg 
canonical or microcanonical) and the gliding average over time $t'$.  

It is useful to split the full stress (for a particular system 
trajectory in the phase space) as a sum of two parts, the stress 
noise, $\sigma_{\alpha\beta}^n(\underline q,t)$, introduced in refs.~\cite{SM2018,Polymers2024}, and 
the deterministic stress, $\sigma_{\alpha\beta}^d(\underline q,t)$, which is equal to the rhs of 
eq.~\ref{ee27} with $\dot{\gamma}$ defined in eq.~\ref{gvr} for 
the very same trajectory:  
\begin{equation}\sigma_{\alpha\beta}(\underline q,t)=\sigma_{\alpha\beta}^n(\underline 
q,t)+\sigma_{\alpha\beta}^d(\underline q,t)\label{snd}\end{equation}
The correlation function of the `stress noise' $\sigma_{\alpha\beta}^n(\underline 
q,t)$  is defined in 
analogy with eq.~\ref{e300}~\cite{Polymers2024,SM2018,SM2025}:  
\begin{equation}C_{\alpha\beta\alpha'\beta'}^n(\underline q,t)\equiv\frac VT\left
<\sigma_{\alpha\beta}^n(\underline q,t+t')\sigma_{\alpha'\beta'}^n(-\underline q
,t')\right>\label{e3n}\end{equation}

The general relation between $C$ and $E$ tensorial functions 
obtained in refs.~\cite{Polymers2024,SM2025} reads:  
\begin{equation}C_{\alpha\beta\alpha'\beta'}(\underline q,s)=E_{\alpha\beta\alpha'
\beta'}(\underline q,s)-\frac {E_{\alpha\beta\nu\mu}(\underline q,s)}{\rho_0s^2}
C_{\mu\delta\alpha'\beta'}(\underline q,s)q_{\delta}q_{\nu}\label{ee48}\end{equation}
where  $C_{\alpha\beta\alpha'\beta'}(\underline q,s)$ is the Laplace-Carson transform of $
C_{\alpha\beta\alpha'\beta'}(\underline q,t)$:  
\begin{equation}C_{\alpha\beta\alpha'\beta'}(\underline q,s)=s\int_0^{\infty}C_{
\alpha\beta\alpha'\beta'}(\underline q,t)e^{-st}{\rm d}t\label{LC}\end{equation}
and $E_{\alpha\beta\alpha'\beta'}(\underline q,s)$ is defined via the same transform.  At $s\to
\infty$ (corresponding to 
$t\to 0$) eq.~\ref{ee48} leads to  
\begin{equation}C_{\alpha\beta\alpha'\beta'}(\underline q,t=0)=E_{\alpha
\beta\alpha'\beta'}(\underline q,t=0)\label{g0}\end{equation}
as can be also obtained directly based on the fluctuation-dissipation 
theorem (FDT)~\cite{SM2018,JCP2022}. 

Eq.~\ref{ee48} can in principle be employed to obtain the elasticity 
tensor $E(t)$ based on the known tensor $C(t)$ of stress correlation 
functions (here we omit both the $\underline q$ variable and the Cartesian 
indices).  This can be done by (i) first obtaining $C(t)$ from 
simulations (or experiments), (ii) then doing its Laplace transform 
numerically, (iii) then solving eq.~\ref{ee48} considered as a matrix 
equation to get $E(s)$ for a given $C(s)$, and (iv) finally doing the 
inverse Laplace transform to obtain $E(t)$.  However, in practice the 
steps (ii) and (iv) are quite problematic since $C(t)$ typically oscillates 
strongly and hence step (ii) requires a lot of data with high sampling 
frequency.  Moreover, the step (iv) of the inverse Laplace 
transform constitutes 
a notoriously difficult and generally ill-posed 
problem~\cite{Phihl,WangZhan}, often requiring a lot of 
computer power. The problem gets even worse  due to a wide time-range sampling requiring to 
characterize the relaxations in glass-forming systems.  

Noteworthily, recently we established another general equation relating the 
elasticity tensor $E$ and the correlation function of stress noise 
$C^n$:~\cite{Polymers2024,SM2018,SM2025} 
\begin{equation}C_{\alpha\beta\gamma\delta}^n(\underline q,t)=E_{\alpha\beta\gamma
\delta}(\underline q,\left|t\right|)-E_{\alpha\beta\gamma\delta}^e(\underline q)= M_{\alpha \beta \gamma \delta}(\underline q,t)\label{f5}\end{equation}
where $E_{\alpha\beta\gamma\delta}^e(\underline q)=E_{\alpha\beta\gamma\delta}(\underline 
q,t\to\infty )$ is the tensor of equilibrium (ie truly 
static) elastic  constants,  and $M$ is the memory kernel of the ZM projection operator formalism \cite{GrimmEtal:SoftMatter2025} ($|t|$ in the rhs is required since $C^\text{n}$ is even in time).
Hence, to obtain the tensor $E$ of generalized 
relaxation moduli using the above relation one has to get the stress 
noise correlation function.  To the best of our knowledge, so far it was impossible  
to measure it experimentally, and, moreover, to get it with MD simulations is very time-consuming
since $\sigma^n$ at time $t$ cannot be directly obtained based on the coordinates 
and momenta of all particles at the same time-point, in contrast to the full stress $\sigma$.  
The way out was 
indicated in ref.~\cite{Polymers2024} and implemented in the velocity-Verlet algorithm as reported in ref.~\cite{BenzeraraSemenovBaschnagel:PRL2026}.  The idea is to suppress the 
deformation rate $\dot{\gamma}_{\alpha'\beta'}(\underline q,t)$ at a given wave-vector $\underline 
q$ (cf  eq.~\ref{ee27}), so that the deterministic stress becomes 
time-independent, $\sigma_{\alpha\beta}^d(\underline q,t)=\sigma_{\alpha\beta}^d
(\underline q)$~\cite{Polymers2024}, hence (cf 
eq.~\ref{snd}) 
\begin{equation}\sigma_{\alpha\beta}(\underline q,t)=\sigma_{\alpha\beta}^n(\underline 
q,t)+\sigma_{\alpha\beta}^d(\underline q)\label{snd0}\end{equation}

Given that the time-averaged stress noise
must vanish:  
$\left<\sigma_{\alpha\beta}^n(\underline q,t)\right>=0$~\cite{Polymers2024},
the full stress correlation function in this case becomes equal to 
\begin{equation}C_{\alpha\beta\alpha'\beta'}^{cmd}(\underline q,t)=C_{\alpha\beta
\alpha'\beta'}^n(\underline q,t)+\frac VT\left<\sigma_{\alpha\beta}^d(\underline 
q)\sigma_{\alpha'\beta'}^d(-\underline q)\right>\label{Cndd}\end{equation}
where the last term, which does not depend on time, was 
identified~\cite{Polymers2024,SM2025} as the equilibrium elasticity 
tensor $E_{\alpha\beta\alpha'\beta'}^e(\underline q)$. Using the basic eq.~\ref{f5} and the above equation 
we then get:
\begin{equation}E_{\alpha\beta\alpha'\beta'}(\underline q,t)=C_{\alpha\beta\alpha'
\beta'}^{cmd}(\underline q,t),~t\geq 0\label{CMD}\end{equation}
where $C_{\alpha\beta\alpha'\beta'}^{cmd}(\underline q,t)$ refers to the stress-correlation tensor for the 
system with the suppressed deformation rate at the given 
wave-vector $\underline q$.  Recalling eq.~\ref{gvr}
we observe that the above constraint is equivalent to ~\cite{Polymers2024,SM2025} 
\begin{equation}\underline J(\underline q,t)\equiv 0\label{e2}\end{equation}
In view of the above discussion we arrive at an important 
conclusion:  Eq.~\ref{CMD} shows that the elasticity tensor $E$ can be 
obtained by calculation of the stress correlation function $C^{cmd}$ of the 
system under the constraint, eq.~\ref{e2}.  
As shown in 
ref.~\cite{Polymers2024}  and \cite{BenzeraraSemenovBaschnagel:PRL2026} the above constraint can be maintained by  an 
appropriate external force field (applied to each particle) 
\begin{equation}\underline F^{ext}_a=m_a (\underline Ae^{{\rm i}\underline q\cdot\underline r_a}+{\rm c}.{\rm c}).\label{e3}\end{equation}
where $\underline A=\underline A(t)$ is complex amplitude of the coherent acceleration, and 
c.c.  means complex conjugate.  The time-dependence of $\underline A$ 
is therefore defined 
by the very condition, eq.~\ref{e2}, and the interaction
forces between the particles.  The molecular dynamics (MD) 
augmented with the external force, eq.~\ref{e3}, such that eq.~\ref{e2} 
is always satisfied, is referred to below as the constrained molecular 
dynamics (CMD).  

 \section{The basic equations of the CMD}

It is worth noting that the condition, eq.~\ref{e2}, does not provide 
any means to {\em directly\/} obtain the coherent acceleration $\underline A(t)$.  In fact, 
once a single wave-vector of interest, $\underline q$, is selected and the 
acceleration $\underline A(t)$ corresponding to this $\underline q$ (cf eq.~\ref{e3}) is 
prescribed, one can find $\underline J(\underline q,t)$ using the standard MD.  However the 
reverse task (to find $\underline A(t)$ based on the prescribed $\underline J(\underline 
q,t)$, which, in the present case, is 
given by eq.~\ref{e2} ) is more difficult. However, in 
practice, this problem can be solved approximately to a sufficiently high precision. 
The implementation of the above ideas in the velocity-Verlet algorithm and its numerical test 
are discussed to some extent in ref.~\cite{BenzeraraSemenovBaschnagel:PRL2026}. Below we extend this discussion by deriving
the approximate basic iteration equations allowing to 
obtain the appropriate time-dependence of the acceleration amplitude 
$\underline A$, and consider their stability.  

The collective displacement of the fluid elements during time $t$ (at the chosen wavevector $\underline q$) is 
defined as 
\begin{equation}\underline U(t)=\int_0^t\underline J(\underline q,t){\rm d}t\label{e4}\end{equation}
where the argument $\underline q$ is omitted for brevity in the lhs (as 
$\underline q$ remains 
constant being pre-selected). As follows from eq.~\ref{f6}, $\underline 
U(t)$ is 
related to $\rho (\underline q,t)$:
\begin{equation}\rho (\underline q,t)=-{\rm i}\rho_0\underline q\cdot\underline 
U(t)+\rho (\underline q,0)\label{e4a}\end{equation}
Eq.~\ref{e2} implies that 
$\underline U(t)$ must be equal to 0 for all times $t \geq 0$, so $\rho(\underline q,t) = \rho(\underline q,0)$ remains constant.

Let us now consider a discrete scheme with $t_s=s\Delta t$, $s=0,1..$, where 
$\Delta t\equiv\tau$ is the MD time-step.  Then eq.~\ref{e4} becomes 
\begin{equation}\underline U_s=\sum_{s'=1}^s\Delta\underline U_{s'}\label{e5}\end{equation}
with 
\begin{equation}\Delta\underline U_s=\int_{t_{s-1}}^{t_s}\underline J(\underline 
q,t){\rm d}t\label{e6}\end{equation}
Imposing that $\underline A(t)$ is constant on each time interval between $t_{s-
1}$ and 
$t_s$:  $\underline A(t)=\underline A_s$ for $t_{s-1}<t\leq t_s$, we conclude that the condition 
$\underline U_s\equiv\underline U(t_s)=0$ for $s=1,2...$ should eventually define the discrete 
values of $\underline A_s\equiv\underline A(t_s)$.  

As mentioned above, any practical numerical scheme based on such a condition is prone to 
be approximate, and therefore some care about its 
stability must always be taken.  In particular, one should identify the region where the 
scheme does not diverge, but rather remains stable. Our preliminary analysis showed that the problem of divergence is severe mostly during the initial stage  when the 
coherent acceleration $\underline A$ is large due to, for example, the initial conditions that strongly violate the constraint, eq.~\ref{e2}.   
Fortunately, in this regime the internal force $F^{int}$ (due to interactions of a particle with its neighbors) can be neglected in comparison with the external force $F^{ext}$, cf 
eq~\ref{e3}. Accordingly, let us assume that 
\begin{equation}F^{int}\ll F^{ext}\label{e6a}\end{equation}
This assumption is considered in more detail and verified in section~\ref{corr}.

Using eq.~\ref{e1} we get (for a single selected wave-vector $\underline q$) 
\begin{equation}\partial\underline J(\underline q,t)/\partial t=\frac 1N\sum_a\mu_
a\left[\partial\underline v_a/\partial t-\underline v_a\left({\rm i}\underline q
\cdot\underline v_a\right)\right]e^{-{\rm i}\underline q\cdot\underline r_a(t)}\label{e7}\end{equation}
where $\mu_a=m_a/\bar {m}$ and
\begin{equation}\partial\underline v_a/\partial t=\left[\underline Ae^{{\rm i}\underline 
q\cdot\underline r_a(t)}+c.c.\right]+\underline F^{int}_a/m_a\label{e8}\end{equation}
Note that the second term in square brackets in eq.~\ref{e7} is 
inertial in nature; it is small and can be neglected as discussed in 
section~\ref{corr}.  The main contribution to $\partial\underline J/\partial t$ therefore comes 
from the first term in square brackets in the rhs of eq.~\ref{e7}, while 
the last term in the rhs of eq.~\ref{e8} can be neglected as well. This leads to the following approximate equation:  
\begin{equation}\partial\underline J/\partial t\simeq\frac 1N\sum_a\mu_a\left[\underline 
Ae^{{\rm i}\underline q\cdot\underline r_a(t)}\right]e^{-{\rm i}\underline q\cdot\underline 
r_a(t)}\label{f9}\end{equation}
It implies coherent contributions from all particles due to 
cancellation of exponential terms leading to 
\begin{equation}\partial\underline J(\underline q,t)/\partial t\simeq\underline 
A(t)\label{e9}\end{equation}
The neglected terms provide minor 
contributions to $\partial\underline J/\partial t$ (see clarifications below and a more 
comprehensive discussion in section~\ref{corr}).  
Note that Eqs.~\ref{e7},~\ref{e8} can be also written as
\begin{equation}\rho_0\partial\underline J(\underline q,t)/\partial t=\underline 
f^{int}+\underline f^{ext}\label{f10}\end{equation}
where $\underline f^{int}$ and $\underline f^{ext}$ are the (Fourier transformed) internal and 
external force densities (ie force per volume), respectively:  
\begin{equation}f^{int}_{\alpha}(\underline q,t)={\rm i}q_{\beta}\sigma_{\alpha\beta}
(\underline q,t)\label{f11}\end{equation}
(cf. eq.~\ref{f7})
\begin{equation}\underline f^{ext}(\underline q,t)=\rho_0\underline A(t)+\rho (2\underline 
q,t)\underline A^{*}(t)\label{f12}\end{equation}
(cf the term in square brackets in eq.~\ref{e8} and 
eqs.~\ref{e7},~\ref{f8}).  Noteworthily, while $\rho (\underline q,t)$ is related to $\underline 
U(t)$, 
cf eq.~\ref{e4a}, the density fluctuation at the doubled wave-vector, 
$\rho (2\underline q,t)$, is independent of both $\underline U(t)$ and the external acceleration $\underline 
A(t)$.  
In other words $\rho (2\underline q,t)$ is independent of the 
constraint, eq.~\ref{e2}, implying that $\underline J(t)=\partial\underline U/\partial 
t=0$.  It is
important that $\rho (2\underline q,t)$ in the above equation, being a fluctuation, 
generally scales as $\rho (2\underline q)\sim\rho_0/\sqrt {N}$ so that the second term in 
eq.~\ref{f12} can be neglected during the initial stage of $J(q,t)$ suppression (cf discussion in 
section~\ref{corr}).  

Furthermore, the internal force $f^{int}_{\alpha}(\underline q,t)$ can be split in two parts: 
\begin{equation}f^{int}_{\alpha}(\underline q,t)=f^{vir}_{\alpha}(\underline q,t
)+f^{id}_{\alpha}(\underline q,t)\label{f13}\end{equation}
where $f^{vir}_{\alpha}(\underline q,t)$ is due to interactions between the particles 
corresponding to the virial part of the stress tensor $\sigma_{\alpha\beta}(\underline 
q,t)$ given 
by the first term in eq.~\ref{G2}.  As for the 
term, $f^{id}_{\alpha}(\underline q,t)$, in eq.~\ref{f13}, it comes (by virtue of eq.~\ref{f11}) 
from the ideal-gas contribution to the stress:  
\begin{equation}\sigma_{\alpha\beta}^{id}(\underline q,t)=-\frac 1V\sum_am_av_{a
\alpha}(t)v_{a\beta}(t)e^{-{\rm i}\underline q\cdot\underline r_a(t)}\label{f14}\end{equation}
(cf the second term in the rhs of eq.~\ref{G2}). 
Note that $f^{id}_{\alpha}(\underline q,t)$ corresponds to the second term 
in square brackets in eq.~\ref{e7}.

  The iterative equations defining the 
time-dependence of $\underline A$, $\underline J$ and $\underline U$ in the main approximation reflected in eq.~\ref{e9} are 
considered in the next two sections.  The effect of the corrections (in particular due to internal forces, $\underline f^{int}$, cf eqs.~\ref{f13},~\ref{f11}) is 
discussed in section~\ref{corr}.

\section{\label{ItEq}Iterative equations and their stability}

The collective displacement, eq.~\ref{e4}, obviously satisfies the following 
differential equation
\begin{equation}\frac {\partial\underline U}{\partial t}=\underline J\label{e10}\end{equation}
Recalling that during each time-step, $t_{s-1}<t\leq t_s$, $(s=1,2...$) the 
coherent acceleration $\underline A=\underline A_s$ is constant, we get from ~eq.~\ref{e9}:
\begin{equation}\underline J_s=\underline J_{s-1}+\underline A_s\tau\label{e11}\end{equation}
Therefore, by virtue of eqs.~\ref{e10},~\ref{e11} 
\begin{equation}\underline U_s=\underline U_{s-1}+\underline J_{s-1}\tau +\underline 
A_s\tau^2/2\label{e12}\end{equation}
Demanding that $\underline U_s=0$ at each step we get: 
\begin{equation}\underline A_s=-\frac 2{\tau^2}\left[\underline U_{s-1}+\underline 
J_{s-1}\tau\right]\label{e13}\end{equation}
Since the condition $U=0$ must hold at any time-step including step $s-1$, we obtain based on eqs.~\ref{e11},~\ref{e13}
\[\underline J_s=-\underline J_{s-1}\]
This means that the collective velocity $J$ does not converge to 0, 
hence the iterative scheme is not really stable.  In an attempt to 
stabilize it, we modify eq.~\ref{e13} as 
\begin{equation}\underline A_s=-\frac {2\kappa}{\tau^2}\left[\underline U_{s-1}+\underline 
J_{s-1}\tau\right]\label{e14}\end{equation}
where $\kappa$ is a constant parameter.  Note that for $\kappa\neq 1$ 
eqs.~\ref{e14},~\ref{e12} yield non-zero $U_s$. 
The following iterative relations for other variables can be derived based on eqs.~\ref{e11},~\ref{e12},~\ref{e14}:  
\begin{equation}\underline J_s=(1-2\kappa )\underline J_{s-1}-2\kappa\underline 
U_{s-1}/\tau\label{e15}\end{equation}
\begin{equation}\underline U_s=(1-\kappa )\left(\underline U_{s-1}+\underline J_{
s-1}\tau\right)\label{e16}\end{equation}
At this point it is natural to expect that stabilization of the recursive 
system of equations described above can be achieved with weaker changes of the 
variables at each step. Therefore we apply the condition $\kappa < 1$ leading to a decrease of the acceleration variable $\underline A$, see eq.~\ref{e14}. Here and below we therefore consider the constant $\kappa$ as the
`stabilization' parameter (see eqs. ~\ref{e14},~\ref{e15},~\ref{e16}).

In Ref.~\cite{BenzeraraSemenovBaschnagel:PRL2026} we implemented this algorithm by introducing a factor $1/\kappa$ in the predictor step (cf.\ Eq.~(11) in \cite{BenzeraraSemenovBaschnagel:PRL2026}), while Eq.~(44) (cf.\ Eq.~(20) in \cite{BenzeraraSemenovBaschnagel:PRL2026}) was applied in the corrector step to calculate the updated positions (Eq.~(21) in \cite{BenzeraraSemenovBaschnagel:PRL2026}) and velocities (Eq.~(22) in \cite{BenzeraraSemenovBaschnagel:PRL2026}) of the particles at the end of the time step.

Considering a certain Cartesian component of the vectors $\vec U_s$ and $\vec J_s$ (for example,  their $x$-components, denoted simply as  $U_s$  and $J_s$ ) and introducing the reduced variable $Y_s\equiv U_s/\tau$  and the formal two-component vector $\vec X_s=(Y_s,J_s)$  we get  from eqs.~\ref{e15},~\ref{e16} :
\begin{equation}\vec X_s=M\cdot \vec X_{s-1}\label{e17}\end{equation}
where 
\begin{equation}M=\left(\begin{array}{cc}
1-\kappa&1-\kappa\\
-2\kappa&1-2\kappa\end{array}
\right)\label{e18}\end{equation}
The eigenvalue equation for the matrix $M$ (with $\lambda$, the eigenvalue) is
\[\left(1-\kappa -\lambda\right)\left(1-2\kappa -\lambda\right)+2\kappa\left(1-\kappa\right
)=0\]
while its general solution reads
\begin{equation}\lambda =1-\frac 32\kappa\pm\sqrt {\kappa\left(9\kappa /4-2\right
)}\label{e19}\end{equation}
The latter equation invites consideration of the following three regimes:

(i) If $\kappa <0$, then
\begin{equation}\lambda =1+1.5\left|\kappa\right|\pm\sqrt {\left|\kappa\right|\left
(2+9\left|\kappa\right|/4\right)}\label{e20}\end{equation}
so that taking the $`+$' option we get the eigenvalue with the largest 
magnitude (among all the roots), $\Lambda\equiv\left|\lambda\right|_{max}$, which exceeds 1:  $
\Lambda >1$.  
Therefore, the iterative process is unstable in this regime.  

(ii) For $0<\kappa <8/9$,
\begin{equation}\lambda =1-1.5\kappa\pm {\rm i}\sqrt {\kappa\left(2-9\kappa /4\right
)}\label{e21}\end{equation}
so that $\Lambda^2=1-\kappa$, leading to stability.

(iii) For $\kappa >8/9$, 
\begin{equation}\lambda =1-1.5\kappa\pm\sqrt {\kappa\left(9\kappa /4-2\right)}\label{g22}\end{equation}
so that the largest magnitude of $\lambda$ (corresponding to the `-' option) is 
$\Lambda =1.5\kappa -1+\sqrt {\kappa\left(9\kappa /4-2\right)}$.  The magnitude $
\Lambda =\left|\lambda\right|_{max}$ therefore 
increases monotonically from $\Lambda =1/3$ at $\kappa =8/9$ to $\Lambda =1$ at $
\kappa =1$, 
and $\Lambda >1$ at $\kappa >1$.  

The minimum of $\Lambda$, $\Lambda_{min}=\min\Lambda (\kappa )$ among real $\kappa$, is therefore attained 
at the optimal $\kappa =\kappa^{*}=8/9$, where the eigenvalue equation has the 
repeated root $\lambda =-1/3$ (hence $\Lambda_{min}=1/3$).  It shows that the general 
solution of eqs.~\ref{e15},~\ref{e16} 
for either of the variables $(\underline A$, $\underline J$ or $\underline U$) has the following form (here we return to vectorial variables):
\begin{equation}\underline J_s=\left(\underline C_0+\underline C_1s\right)\left(
-1/3\right)^s\mbox{\rm \  for }\kappa =\kappa^{*}=8/9\label{g23}\end{equation}
where $\underline C_0$, $\underline C_1$ are constant (and generally complex) vectors, which 
depend on the initial conditions:
\begin{equation}\underline U_0=0,~\underline A_0=0,\mbox{\rm \ and arbitrary }\underline 
J_0\label{g24}\end{equation}
Note that the first condition above follows from eq.~\ref{e4}.
Eq.~\ref{g23} shows that $\underline J$ rapidly vanish in time, and the same is 
true for other variables.  Using eqs.~\ref{g23},~\ref{e15},~\ref{e16} we 
then get (again for $\kappa^{*}=8/9$), see Appendix~\ref{AppB}:  
\begin{equation}\underline J_s=-\underline J_0\left(3+4s\right)\left(-1/3\right)^{
s+1},~\underline U_s=\tau\underline J_0s\left(-1/3\right)^{s+1}\mbox{\rm \ , }s=
0,1,2..\label{g25}\end{equation}

The algorithm described above was employed in Ref.~\cite{BenzeraraSemenovBaschnagel:PRL2026} to provide a proof-of-principle for the CMD method. In this case by virtue of 
eq.~\ref{e4a} the density fluctuation $\rho (\underline q,t)$ is kept almost exactly constant 
and equal to the initial value $\rho (\underline q,0)$. 
Instead of eq.~\ref{g24}, one may choose to employ another initial condition for $\underline U_0$, which works to suppress the density fluctuation $\rho(\underline q,t)$ forcing it to approach zero after several MD time steps $\tau$. To describe this option it is useful to consider ``naturally rotated coordinates'' (NRC) \cite{WittmerEtal_P5a:PRE2023} where the axis `1' is defined by the unitvector $\widehat{e}_1 = \underline{q} / q$ parallel to the selected wavevector $\underline q$, while the other axes, with unitvector $\widehat{e}_2$ in two dimensions (2D) and unitvectors $\widehat{e}_2, \widehat{e}_3$ in three dimensions (3D), are chosen 
such that a right-handed basis is obtained. 
The difference between eq.~\ref{g24} and this alternative choice of initial conditions 
concerns only the longitudinal components of the vectors  $\underline U$, $\underline 
J$ and $\underline A$:  
$U^L=\underline U\cdot\underline q/q$, etc.  Namely, we define $U^{wL}\equiv U^L+{\rm i}
\rho (\underline q,0)/\left(\rho_0q\right)$ and 
employ the same eqs.~\ref{e14},~\ref{e15},~\ref{e16}, where $\underline A$ is 
replaced with $A^L$, $\underline J$ with $J^L$ and $\underline U$ with $U^{wL}$.  Since $U^L_0=0$, the 
initial condition for $U^{wL}$ is $U_0^{wL}={\rm i}\rho (\underline q,0)/\left(\rho_
0q\right)$ and, moreover, at 
further steps it remains proportional to $\rho (\underline q,t)$:
\begin{equation}U^{wL}_s={\rm i}\rho (\underline q,t_s)/\left(\rho_0q\right)\label{g26}\end{equation}
due to eq.~\ref{e4a}.  By the nature of 
eqs.~\ref{e14},~\ref{e15},~\ref{e16} discussed above, all variables 
including $U^{wL}$ must tend to zero as a function of time for $0<\kappa <1$, and this means that 
the present algorithm tends to keep the relevant density 
fluctuation $\rho (\underline q,t)$ equal to zero at $t\gg\tau$.

 Thus, the only essential difference between the two algorithms (associated with eqs.~\ref{g24} and~\ref{g26} concerns 
 the initial condition for the longitudinal component of the collective displacement: $U^{L}$ vs. $U^{wL}$. 
 To summarize, the second option implies the following initial conditions (in 3D):
	\begin{equation}
		U_0^\text{wL} = {\rm i}\frac{\rho(\underline q,0)}{\rho_0 q}, \; U_0^{(2)} = U_0^{(3)} = 0, \quad \underline{A}_0 = 0, \quad \mbox{and arbitrary} \;
		\underline{J}_0 ,
		\label{eq:IC2}
	\end{equation} 
	where $U_0^{(2)}$ and $U_0^{(3)}$ denote respectively the components of $\underline{U}$ 
    along $\widehat{e}_2$ and $\widehat{e}_3$. (Note that in practice the initial current $\underline{J}_0$
is defined by the initial micro-state of the molecular system.

\section{\label{Alt}Alternative approach}

The results of the previous section can be compared with the stability behavior of the 
basic variables ($J$, $U$, $A$) coming from iterative equations based on a 
different approach to impart the stability.  
The alternative approach  involves the following steps:  

(i) First we assume that $\vec A_s=\vec A_{s-1}$ (recall that we still assume that 
$A(t)=A_s={\rm c}{\rm o}{\rm n}{\rm s}{\rm t}\,$ in the time interval $t_{s-1}<t
\leq t_s$) and find $J_s$ and $U_s$ 
using eqs.~\ref{e9},~\ref{e10}.  In particular, this implies that $\underline U$ at time $t_s$ is predicted to be $\underline{U}^\text{p}_s = \underline{U}_{s-1} + \underline{J}_{s-1} \tau + \underline{A}_{s-1} \tau^2 / 2$.

(ii)  Then, we note that the resulting $\underline{U}^\text{p}_s$ is generally nonzero. 
To impose that the  collective displacement is vanishing it is necessary to change the external force appropriately. This can be achieved by introducing a correction $\widetilde{\underline A}_s$ to the coherent acceleration, $A_s$:
\begin{equation}\vec A_s=\vec A_{s-1}+\tilde {\vec A}_s\label{e24}\end{equation}
with $\tilde {\vec A}$$_s$ defined by the condition $\vec U_s=0$ leading to 
\begin{equation}\vec{\tilde {A}}_s=-\frac 2{\tau^2}\left[U_{s-1}+J_{s-1}\tau +A_{s-1}\tau^
2/2\right]\label{e25}\end{equation}
Since eqs.~\ref{e24}, ~\ref{e25} give back eq.~\ref{e13},  they lead to marginal stability, as in the previous section. To stabilize the iterative process we follow the approach of Sec.~4 and modify eq.~\ref{e25} by reducing the `correction term' $\widetilde{\underline A}_s$:  
\begin{equation}\vec{\tilde {A}}_s=-\frac {2\kappa}{\tau^2}\left[\vec U_{s-1}+\vec J_{s-1}\tau +
\vec A_{s-1}\tau^2/2\right]\label{e26}\end{equation}
with $\kappa <1$. As a result we get
\begin{equation}\vec U_s=(1-\kappa )\left[\vec U_{s-1}+\vec J_{s-1}\tau +\vec A_{s-1}\tau^2/2\right]\label{e27}\end{equation}
\begin{equation}\vec J_s=(1-2\kappa )\vec J_{s-1}+(1-\kappa )\vec A_{s-1}\tau -\frac {2\kappa}{
\tau}\vec U_{s-1}\label{e28}\end{equation}
\begin{equation}\vec A_s=(1-\kappa )\vec A_{s-1}-\frac {2\kappa}{\tau}\left[\vec J_{s-1}+\vec U_{s-1}
/\tau\right]\label{e29}\end{equation}

Thus we arrive at a system of 3 recursive equations for the collective 
acceleration $\vec A$, the collective velocity $\vec J$ and the collective 
displacement $\vec U$ (whose gradient corresponds to the collective strain, 
$\gamma_{\alpha\beta}={\rm i}U_{\alpha}q_{\beta}$, at the selected wavevector $\underline 
q$).  

Now considering , say, $x$-components of all the three variables, $A$, $J$ and $U$, as a formal vector 
\begin{equation}\vec{X}=\left(A\tau^2/2,~J\tau ,~U\right)\label{e30}\end{equation}
we get based on eqs.~\ref{e27},~\ref{e28},~\ref{e29} the master recursive 
equation in the vectorial form:  
\begin{equation}\vec{X}_{s}=M \cdot \vec{X}_{s-1}\label{e31}\end{equation}
where now $M$ is a $3\times 3$ matrix
\begin{equation}M=\left(\begin{array}{ccc}
1-\kappa&-\kappa&-\kappa\\
2(1-\kappa )&1-2\kappa&-2\kappa\\
1-\kappa&1-\kappa&1-\kappa\end{array}
\right)\label{e32}\end{equation}
The eigenvalue equation corresponding to eqs.~\ref{e31},~\ref{e32}, 
${\rm d}{\rm e}{\rm t}(M-\lambda )=0$, leads to
\begin{equation}P(\lambda ,\kappa )\equiv\lambda^3-\left(3-4\kappa\right)\lambda^
2+(3\lambda -1)\left(1-\kappa\right)=0\label{e33}\end{equation}
For $\kappa =1$ it can be factorized as
\[\left(\lambda +1\right)\lambda^2=0\]
with obvious roots $\lambda =0,~0,~-1$ pointing to marginal stability. For 
$\kappa =2/3$ the equation becomes 
\[\left(\lambda -1/3\right)\left(\lambda^2+1\right)=0\]
giving 3 distinct eigenvalues, $\lambda =1/3,~{\rm i},~-{\rm i}$, which again corresponds 
to marginal stability.  Finally, with $\kappa =0$, the 
eigenvalue equation reduces to 
\[\left(\lambda -1\right)^3=0\]
Thus, all the three cases considered above correspond to marginal stability.  

A simple numerical analysis of eq.~\ref{e33} shows that the most 
unstable mode corresponding to the largest eigenvalue (with maximum 
$\left|\lambda\right|$ among all roots), $\Lambda =\left|\lambda\right|_{max}$, changes with the parameter $
\kappa$ in 
the following way:  $\Lambda =1$ at $\kappa$ equal to $0$, 2/3 and 1, $\Lambda >
1$ for 
$0<\lambda <2/3$, and $\Lambda <1$ at $2/3<\kappa <1$.  As mentioned in the previous 
section the time-dependencies of all variables $(A$, $J$, $U$) are expected 
to asymptotically follow the $\Lambda^s$ law for large $s$, so that the most 
stable behavior coming from the recursive equations corresponds to 
the minimum of $\Lambda (\kappa )$ equal to $\Lambda_{min}=\Lambda (\kappa^{*})$.  It is natural to expect 
that the optimal conditions correspond to the point where the 
characteristic polynomial $P(\lambda ,\kappa )$ considered as a function of $\lambda$ (cf.  eq.~\ref{e33})  
shows a repeated knot (like, say, $\lambda_
1=\lambda_2$) :  at such a point both roots 
(say, $\lambda_1=\lambda_1(\kappa )$ and $\lambda_2(\kappa )$, defined by equation $
P(\lambda ,\kappa )=0$) generically 
produce singularities which result in kinks for the functions $\left|\lambda_1 (\kappa)\right|$ or $\left
|\lambda_2 (\kappa)\right|$.  In fact, it is 
straightforward to see that such kinks are indeed implied by 
eqs.~\ref{e21},~\ref{g22} considered in the previous section.  
Therefore, the optimal conditions are likely to be defined by the following equations satisfied by the repeated root:  
\begin{equation}P(\lambda ,\kappa )=0,~\partial P(\lambda ,\kappa )/\partial\lambda 
=0\label{e34}\end{equation}
which lead to  
\[\lambda =-1/2,~\kappa =\kappa^{*}=27/28,~\Lambda_{min} =1/2~\]
The three eigenvalues at $\kappa^{*}$ defined by the equation $P(\lambda ,\kappa^{
*})=0$ are 
\begin{equation}\lambda_1=\lambda_2=-1/2,~\lambda_0=1/7\label{e35}\end{equation}
We checked that $\Lambda (\kappa )>\Lambda_{min}$ for any real $\kappa\neq\kappa^{*}$. A rigorous 
way to obtain the above results is outlined in the Appendix~\ref{AppA}.  

It is thus found that the minimum of $\Lambda$, $\Lambda_{min}=0.5$, does correspond to 
$\kappa^{*}=27/28$.  This choice of the parameter $\kappa$ is therefore proved to be 
optimal for the dynamics defined by 
eqs.~\ref{e27},~\ref{e28},~\ref{e29}.  Recalling that $\Lambda_{min}=1/3$ for the 
stabilized dynamics corresponding to eqs.~\ref{e14},~\ref{e15},~\ref{e16}, 
we conclude that the latter stabilization approach is superior as 
compared to the one implied by eqs.~\ref{e27},~\ref{e28},~\ref{e29} considered in the present section.

\section{\label{corr}The effect of subdominant terms and conditions 
for negligible internal forces} 

The analysis of the previous sections related to implementation of 
the condition~\ref{e2} was based on two main equations: the exact 
eq.~\ref{e4} and approximate eq.~\ref{e9}. Restoring the neglected terms 
using eqs.~\ref{f10},~\ref{f11},~\ref{f12} we get the exact equation 
generalizing eq.~\ref{e9}:
\begin{equation}\partial J_{\alpha}(\underline q,t)/\partial t=A_{\alpha}(t)+\frac {
\rho (2\underline q,t)}{\rho_0}A^{*}_{\alpha}(t)+\frac {{\rm i}q_{\beta}}{\rho_0}
\sigma_{\alpha\beta}(\underline q,t)\label{e36}\end{equation}
The absolute value of the ratio of the second term to the 
first term in the above equation is $\frac {\left|\rho (2\underline q,t)\right|}{
\rho_0}$. The variance of this 
ratio is (note that $\left<\rho (\underline q,t)\right>=0$ for $q\neq 0$)
\begin{equation}\mbox{var}\left|\rho (2\underline q)/\rho_0\right|=S_2(2q)/N\label{f15}\end{equation}
where $S_2(q)$ is the mass-weighted structure factor \cite{BaschnagelSemenov:SoftMatter2026}:
	\begin{equation*}
		S_2(q) = \frac{1}{N} \sum_{a,b=1}^N \frac{m_a m_b}{\overline{m}^2}
		\left \langle \exp \left ( \text{i} \underline{q} \cdot (\underline{r}_a - \underline{r}_b )\right )\right \rangle \label{f16}
	\end{equation*}
	For equal masses ($m_a = m$), the above equation agrees with the definition of the standard structure factor $S(q)$~\cite{SM2021p7867,VogelFuchs:EPJE2020} ,
	\begin{equation}
	    S(q) = \frac{1}{N} \sum_{a,b=1}^N 
	    \left \langle \exp \left ( \text{i} \underline{q} \cdot (\underline{r}_a - \underline{r}_b )\right )\right \rangle \label{Bf16}	
	\end{equation}
	To obtain a numerical estimate for the second and third terms of eq.~\ref{e36}, we will use a glass-forming two-dimensional (2D) polydisperse Lennard--Jones (pLJ) model \cite{TanguyEtal:PRB2002}, which we have studied extensively in the past \cite{WittmerEtal:JCP2013,WittmerEtal_P5a:PRE2023,BenzeraraSemenovBaschnagel:PRL2026}.
	
	The 2D pLJ model is defined as follows: Two particles $a$ and $b$, with diameter $\sigma_a$ and $\sigma_b$, interact by the potential $U(r) = 4 \epsilon [(\sigma_{ab}/r)^{12} - (\sigma_{ab}/r)^{6}]$, where the energy parameter $\epsilon$ is the same for all particles, while $\sigma_{ab}$ is defined by the Lorentz mixing rule $\sigma_{ab} = (\sigma_a + \sigma_b) / 2$ \cite{HansenMcDonald}. The potential is truncated and shifted to zero at $r_\text{cut} = 2 r_\text{min} = 2 \times 2^{1/6} \sigma_{ab}$ with $r_\text{min}$ being the distance at which $U(r)$ is minimum. The particle diameters are uniformly distributed between $0.8 \overline{\sigma}$ and $1.2 \overline{\sigma}$ around the average diameter $\overline{\sigma}$, leading to a size polydispersity index of $\overline{\sigma^2} / \overline{\sigma}^2 - 1 \approx 0.013$. All particles have the same mass $m$. In the following, all quantities are reported in LJ units, i.e.\ distance in units of $\overline{\sigma}$, energy in units of $\epsilon$ and temperature in units of $\epsilon/k_\text{B}$. For the numerical estimates, we will use results for an equilibrated 2D pLJ system, which were obtained from molecular dynamics (MD) simulations at $T=0.4$ (which is larger than the glass transition temperature $\Tg \approx 0.26$ \cite{ KlochkoEtal:SoftMatter2021,WittmerEtal:JCP2013}). The system contains $N = 10^4$ particles in a volume $V(T) = 11283.54$ corresponding to the pressure $p = 2$. This volume implies that the number density of particles is $c_0 = N / V \approx 0.886$ and that the norm of the smallest wave-vector, which is compatible with the periodic boundary conditions of the MD simulation, has the value $q_\text{min} = 2 \pi / \sqrt{V} \approx 0.06$. Since all particles have equal mass, we have $\rho_0 = \overline{m} c_0 = c_0$ and $S_2(q) = S(q)$. For $q_\text{min}$ the structure factor is small, $S(q_\text{min}) \approx 0.055$, as is typical of liquids. The maximum of $S(q)$ is found at $q_\text{max} \approx 6.32$, where $S_\text{max} = S(q_\text{max}) \approx 3.769$ \cite{KlochkoEtal:SoftMatter2021}.
	
	Using these numerical values the relative correction due to the $A^*$-term in eq.~\ref{e36} can be estimated from eq.~\ref{f15}:   	\begin{equation}
			\text{For}~ 2 q \sim q_{\text{min}}:
				\Delta _A = \operatorname{rms} \left | \rho(2q) / \rho_0 \right | \sim 0.235 / \sqrt{N} , \label{f16a}
			\end{equation}  \begin{equation}  
			\\ \text{For}~ 2 q \sim q_{\text{max}}: 
			\Delta _A = \operatorname{rms} \left | \rho(2q) / \rho_0 \right | \sim 2 / \sqrt{N} \label{f16b}
			\end{equation}
	Here "rms" is  shortcut for "root-mean-square". We see that the correction is quite small for $N \gtrsim 10^4$. Therefore, the $A^*$-term in eqs.~\ref{e36} and~\ref{f12} can be safely neglected in most cases, in particular, for low $q$ (and large $N$). (Recall that  for the 2D pLJ system we studied
the main peak value of $S_2(q)=S(q)$ is typically $\sim 4$~\cite{SM2021p7867}.)  

Let us turn to the last stochastic term in eq.~\ref{e36}, 
$X_{\alpha}=\frac {{\rm i}q_{\beta}}{\rho_0}\sigma_{\alpha\beta}(\underline q,t)$.  Its mean-square is related to the stress 
correlation function (cf eq.~\ref{e300}):  
\begin{equation}\left<X_{\alpha}X_{\alpha}^{*}\right>=\frac {q_{\beta}q_{\beta'}}{
\rho_0^2}\frac TVC_{\alpha\beta\alpha\beta'}(\underline q,t=0)\label{f17}\end{equation}
By virtue of eq.~\ref{g0} the above equation leads to
\begin{equation}\left<X_{\alpha}X_{\alpha}^{*}\right>=\frac {q_{\beta}q_{\beta'}}{
\rho_0^2}\frac TVE_{\alpha\beta\alpha\beta'}(\underline q,t=0)\label{f18}\end{equation}
In what follows we focus on the longitudinal components of $\underline X$ , $X_1\equiv X_L$
(corresponding to the index $\alpha =1$ in the NRC , ie\ to the component parallel to $\underline{q}$)  since such 
components are more important as compared with the  transverse components which are 
normally weaker.  Then eq.~\ref{f18} simplifies as 
\begin{equation}\left<X_LX_L^{*}\right>=\frac {q^2}{\rho_0^2}\frac TVL_A(\underline 
q)\label{f19}\end{equation}
where $L_A(\underline q)\equiv E_{1111}(\underline q,t=0)$ is the so-called {\em affine\/} longitudinal modulus 
(providing an instantaneous stress response to a weak longitudinal 
strain).  The correction due to the last term $X_L$ is therefore  ~ 
\begin{equation}\mbox{ rms}\left|X_L\right|=(qT/\bar {m})N^{-1/2}\sqrt {L_A(\underline 
q)/\left(Tc_0\right)}\label{f20}\end{equation}
where $c_0=N/V$.  The relative correction due to internal forces can 
be then estimated as 
\begin{equation}\Delta_{int}=\mbox{ rms}\left|X_L\right|/\mbox{\rm rms}\left|
A_L\right|\label{f21}\end{equation}
To quantify it  for the 2D pLJ system we use the numerical values considered below eq.~\ref{Bf16}  ($N=10^4$, 
$\bar {m}=1$, box size $L_b\sim 100$, $q=q_{min}=2\pi /L_b\sim 0.06$, $c_0\sim 1$, ~ref.~\cite{JCP2022}),  noting  that $L_\text{A} (q_\text{min}) \sim 90$ for $T = 0.4$  in LJ units~\cite{JCP2022,SemenovBaschnagel:Polymers2024}.  (Note also
 that $L_\text{A} (q_\text{min})$ is an upper bound because  $L_\text{A} (q_\text{min}) > L_\text{A} (q > q_\text{min})$~\cite{SemenovBaschnagel:Polymers2024,kstress}.) 
  Thus we get using eq.~\ref{f20} for 
$T=0.4$:  
\begin{equation}\mbox{rms}\left|X_L\right|\sim 4\cdot 10^{-3}\label{f21a}\end{equation}
 To estimate $|A_L|$ for the initial conditions of eq.~(54), we argue that $|A_L| \sim |J_L| / \tau$,  since the external acceleration should work to nearly 
compensate for the initial collective velocity $\underline J(\underline 
q,t=0)$.  
Recalling that the typical thermal amplitude of the collective velocity 
is $\left|J_L\right|\sim v_T/\sqrt {N}$ (as follows from the fact that initial velocities of 
all particles are independent and therefore uncorrelated, and the 
definition, eq.~\ref{e1}), where $v_T=\sqrt {T/\bar {m}}$ is the typical thermal 
velocity of a particle, we find 
\begin{equation}\mbox{\rm rms}\left|A_L\right|\sim v_T/\left(\tau\sqrt {N}\right
)\label{f22a}\end{equation}
For $\tau =0.005$ it leads to 
\begin{equation}\mbox{\rm rms}\left|A_L\right|\sim 1\label{f22}\end{equation}
so that the relative internal correction (cf. eq.~\ref{f21}) is 
\begin{equation}\Delta_{int}\sim 3\cdot 10^{-3}\label{f23}\end{equation}
The relative correction $\Delta_A$ due to the $A^{*}$-term in eq.~\ref{e36} is a 
bit smaller being $\Delta_A\sim 0.25\%$ (cf eq.~\ref{f16a}) hence the total 
correction is $\Delta\sim 0.4\%$, see eq.~\ref{f24} below.  (Note that the second 
and third terms in eq.~\ref{e36} are virtually uncorrelated as long as 
$\underline A$ is considered as an independent variable, which is plausible in the 
case of eq.~\ref{g24} since $\underline A$ is virtually proportional to $\underline J(\underline 
q,t=0)$ and therefore is essentially uncorrelated to $\sigma (q,t=0)$, hence in practice it is the variances of the two 
corrections that should be added to obtain the total correction.)  
Thus, the total relative correction to the main approximation, 
eq.~\ref{e9}, is 
\begin{equation}\Delta =\sqrt {\Delta_A^2+\Delta_{int}^2}\label{f24}\end{equation}
with 
\begin{equation}\Delta_A\sim\sqrt {S_2(2q)/N},~~\Delta_{int}\sim q\tau\sqrt {\frac {
L_A(q)}{c_0\bar {m}}}\label{f25}\end{equation}
(cf eqs.~\ref{f15},~\ref{f16a},~\ref{f20},~\ref{f21},~\ref{f22a}).

Let us now turn to the initial conditions associated with eq.~\ref{g26} where the external 
acceleration should compensate for the initial collective longitudinal 
displacement $U^{wL}_0={\rm i}\rho (\underline q,0)/\left(\rho_0q\right)$ (cf eq.~\ref{g26}), leading to 
$\left|A_L\right|\tau^2\sim\left|\rho (\underline q,0)\right|/\left(\rho_0q\right
)$.  With 
$\left|\rho (\underline q,0)\right|/\rho_0\sim\left[S_2(q)/N\right]^{1/2}\sim\left
[0.06/10^4\right]^{1/2}$ (cf eq.~\ref{f15}) we thus 
arrive at $\left|A_L\right|\sim 1600$  giving rise to the relative internal correction 
\begin{equation}\Delta_{int}\sim 2.5\cdot 10^{-6}\label{f26}\end{equation}
which is totally negligible.  The correction $\Delta_A\sim 2.5\cdot 10^{-3}$ (cf 
eq.~\ref{f16a}) dominates in this case.  

Since the arguments presented in Sec.~3 hinge on neglecting the internal forces, $F^\text{int} \ll F^\text{ext}$ [cf.~eq.~\ref{e6a}], it is also of interest to consider similar corrections for two `idealized' systems: 
	
	First, an ideal-gas system subjected to the initial conditions of eq.~\ref{g24}. In this case, $S(q)=1$ and $L_A(q)=3c_0T$~\cite{JCP2022}, hence (cf eqs.~\ref{f25})
\begin{equation}\Delta_A\sim 1/\sqrt {N},~~\Delta_{int}\sim\sqrt {3}qv_T\tau\label{f27}\end{equation}
Using the same parameters as before we thus get 
$\Delta_A\sim 1\%$,~$\Delta_{int}\sim 0.03^{\%}$.  The $\Delta_A$ correction is therefore most relevant here again.
	
Second, an ideal gas system whose initial configuration (at $t = 0$)   
is taken from a well-equilibrated system of 
interacting pLJ particles. In this case the corrections are given by 
eqs.~\ref{f25} with $S_2(2q)$ for the interacting particles  and [cf.~eqs.~\ref{f18},~\ref{f17}] 
	\begin{equation}
		L_A(q) = E_{1111}(q,t=0) = C_{1111}(q,t=0) = c_0 T \left ( 2 + S(q) \right )\label{f28}
	\end{equation}
where the last equality is obtained by inserting the kinetic part of the stress tensor [i.e.\ the second term of eq.~\ref{G2}] into eq.~\ref{e300} and noting that the particle velocities are Gaussian-distributed at equilibrium.
 Eq.~\ref{f28}  reflects the dependence of the affine modulus $L_A$ 
on the system configuration.  It can be obtained using the definition 
of $L_A(q)=L(q,0)$, cf ref.~\cite{Polymers2024}.  It shows that 
 $L_A$ is roughly the same for the fully equilibrated ideal gas and in 
the case of artificial initial configuration, hence the estimates given 
in eqs.~\ref{f27} are applicable also in the artificial case.  

In the four figures below (Fig. 1 to Fig. 4) we compare the theoretical results based on 
the approximate recursive equations (cf section~\ref{ItEq}) with 
the simulation data for the 2D pLJ model at $T=0.4$, which have been obtained from an implementation of the CMD approach in the velocity-Verlet algorithm as explained in~\cite{BenzeraraSemenovBaschnagel:PRL2026}. The simulation studies are based on several initial microstates taken from equilibrium ensemble. Two of them, labeled as configuration 3 and 4, are used to produce the figures. 
The variables shown in the graphs are reduced 
according to the following rules:  $\mathcal A=AN\tau^2/2$, $\mathcal J=JN\tau$, 
$\mathcal U=UN$, and $\mathcal U^{wL}=U^{wL}N$, where the physical variables $A$ (collective acceleration at wave-vector $\vec{q}$), $J$ (collective velocity) and $U$ (collective displacement) are defined in 
the previous sections.  The reduced variables are shown as a 
function of the MD step $s$.  The reduced {\em theoretical\/} data are shown by small squares.
The title like `Re(Transverse)' above a graph 
means that it shows real parts of transverse variables, while, say, 
`Im(Longitudinal)' points to imaginary parts of longitudinal variables.  
The comparison between simulation and theory is presented for both initial conditions, 
eq.~\ref{g24} (Figs.~1--2) and another condition associated with eq.~\ref{g26} (Figs.~3--4).

An excellent agreement between theory and simulation data can be 
observed.  It shows that the internal forces are indeed negligible 
during the transient stabilization stage in all the systems we studied 
by MD simulations. However, while  the time-dependencies of the acceleration 
$\underline A(t)$, the collective velocity $\underline 
J(t)$ and the collective 
displacement $\underline U(t)$ obtained by simulations very closely follow the theoretical 
predictions (described in section~\ref{ItEq}) for the first 5--10 MD steps, the relative deviations start to grow at longer times due to exponential decrease of the tracked variables leading to  stronger {\sl relative} corrections coming from internal forces and inertial effects.

\section{Summary}

In this paper we elucidated how the constraint of no flow at
selected wave-vector $\underline q$ (cf eq.~\ref{e2}) can be implemented in the 
molecular dynamics algorithm as applied to fluids, glass-forming 
liquids and amorphous solids.  In particular, we derived approximate 
recursive equations defining the time-dependencies of the external acceleration 
amplitude $\underline A(t)$, the collective velocity $\underline J(t)$ and displacement $\underline 
U(t)$, 
all at the selected wavevector $\underline q$, during the transient stage of the system 
relaxation from the initial  equilibrium state (at $t=0$) to the new state of conditional equilibrium under the constraint, eq.~\ref{e2}, imposed at 
$t>0$.  Considering several different stabilization techniques (cf sections~\ref{ItEq} and~\ref{Alt}), we show that the constraint, eq.~\ref{e2}, can be accurately maintained leading to the steady (partially equilibrium) state which is readily  
achieved at $t\gg\tau$.  The most efficient approach is considered in section~\ref{ItEq}. It 
shows that the fastest decay of the collective current $\underline J(t)$ is achieved at the predicted optimal value of the stabilization parameter, $\kappa=8/9$. In this case both $\underline J(t)$ and the external force proportional to $\underline A(t)$ show a swift convergence to nearly zero following an 
exponential decay, $J,A\propto 1/3^s$ (cf eq.~\ref{g23}), with small 
fluctuations of relative order $10^{-3}$ or lower in the steady state  
(here $s=t/\tau$ is the number of MD time-steps).  

It is also demonstrated that the transient stage typically takes about 10 
MD steps.  It is important to emphasize that in the transient regime 
the internal forces on the particles (cf eq.~\ref{f11}) can be 
neglected and that is why the recursive equations are simple.  
The theoretical results for the time-dependence of the main variables, $\underline J(t)$, $\underline U(t)$, $\underline A(t)$, during the transient stage show an excellent agreement with the simulation data, as shown in Figs.~1--4. It is also remarkable that the relaxation processes shown in Figs.~3  for the perfect gas and in Fig.~4 for the pLJ system of interacting particles (with the same initial microstate) are nearly coinciding during the transient stage. This observation underlines our conclusion that particle interactions are irrelevant in the transient regime.

The contribution of subdominant terms to the exact dynamical equation (eq.~\ref{e36})
at the beginning of the 
transient stage is estimated in eqs.~\ref{f25} for the stabilization 
algorithm corresponding to eq.~\ref{g24}.  The first correction $\Delta_A$ (which is related to density fluctuations) is proportional to $1/\sqrt N$ (with $N$, 
the number of particles in the system); this correction is therefore vanishing in the thermodynamic limit.  The second correction related to pairwise interactions, $\Delta_{int}$, 
decreases as the MD time-step $\tau$ gets shorter, but, on the other hand, it increases if the wavevector $q$ is increased (cf eqs.~\ref{f25}).  
Note however that the $q$-dependence of $\Delta_{int}$ is expected to be weaker than linear due to a decrease of the instantaneous (affine) elastic modulus $L_A(q)$ with $q$, as follows from a recent analysis~\cite{kstress}. 

It is also worth noting that the stabilization algorithm associated with eq.~\ref{g26} leads to much stronger variations of longitudinal components of 
$\underline J(t)$, $\underline U(t)$, $\underline A(t)$ than the similar algorithm with initial conditions specified in eq.~\ref{g24}
(compare the bottom graphs in Fig. 2 and Fig. 4). The difference amounts to a factor of 300 or more. This strong effect is rooted in different targets of the two algorithms: the first one (associated with eq.~\ref{g24}) is aimed at keeping $\underline J(t)$ as small as possible, while the second one is targeted, in addition, at suppressing also the density fluctuation $\rho(q,t)$.

\newpage

\appendix

\section{\label{AppB} Derivation of eq.~\ref{g25}}

For the eigenvalues [see eq.~\ref{e19}]
	\begin{equation}
		\lambda_\pm = 1 - \frac 32 \kappa \pm \sqrt{\frac{\kappa}{4}(9\kappa -8)} ,
		\label{eq:evalueApp}
	\end{equation}
	the eigenvectors $\underline{u}_\pm = (u_{1,\pm},u_{2,\pm})$, obtained from $M \cdot \underline{u}_\pm = \lambda_\pm \underline{u}_\pm$, can be chosen as
	\begin{equation}
		u_{1,\pm} = 1, \quad
		u_{2,\pm} = \frac{-\kappa \pm \sqrt{\kappa (9 \kappa - 8)}}{2(1 - \kappa)} .
		\label{eq:evpm}
	\end{equation}
	The optimal $\kappa = \kappa^* = 8/9$ implies that the eigenvalue equation~\eqref{eq:evalueApp} has the repeated root $\lambda_+ = \lambda_- = \lambda = - 1/3$, leading to one eigenvector only
	\begin{equation}
        \underline{u}_+ = \underline{u}_- = \underline{u} =
        \bigg (
        \begin{array}{rr}
        	1 \\[-1.5ex] 
        	-4
        \end{array}
        \bigg ) .
        \label{eq:evu} 
	\end{equation}
	Using $\underline u$, a further (generalized) eigenvector $\underline v = (v_1,v_2)$ can be defined by
	\begin{equation}
		(M - \lambda I)\cdot \underline v = \underline u
		\quad \Rightarrow \quad
		M \cdot \underline v = - \frac 13 \underline v + \underline u ,
		\label{eq:gev}
	\end{equation}
	where $I$ is the identity matrix and $M$ is given by eq.~\ref{e18} with $\kappa = 8/9$. A solution of eq.~\eqref{eq:gev} is
	\begin{equation}
		\underline v =
	    \bigg (
		\begin{array}{cc}
			v_1 \\[-1.5ex] 
			9 - 4 v_1
		\end{array}
		\bigg )
		= 
		\bigg (
		\begin{array}{cc}
			0 \\[-1.5ex] 
			9
		\end{array}
		\bigg )
		+ v_1 \underline u .
        \label{eq:gev1}
	\end{equation}
	Next, we consider the vector $\underline X_s = (Y_s,J_s)$ and write its initial condition as
	\begin{equation}
		\underline{X}_0 = 
				\bigg (
		\begin{array}{cc}
			Y_0 \\[-1.5ex] 
			J_0
		\end{array}
		\bigg )
		= C_0 \underline u + C_1 \underline v ,
		\label{eq:X0}
	\end{equation}
	where $C_0$ and $C_1$ are constants. Iterating eq.~(47) leads to $\underline{X}_s = M^s \cdot \underline{X}_0$. So, we need $M^s \cdot \underline{v}$ and $M^s \cdot \underline{u}$, due to eq.~\eqref{eq:X0}. Since $M^s \underline{u} = (-1/3)^s \underline{u}$, iteration of eq.~\eqref{eq:gev} yields
	\begin{equation}
		M^s \cdot \underline{v} = \left ( - \frac 13\right )^s \underline{v} + s \left ( - \frac 13\right )^{s-1} \underline{u} . 
		\label{eq:Msv}
	\end{equation}
	Using eq.~\eqref{eq:gev1} we then obtain
	\begin{equation}
		\underline{X}_s = \left ( - \frac 13\right )^s \bigg  [ 
		(C_0 + v_1) \underline u + C_1 
		\bigg (
		\begin{array}{cc}
			0 \\[-1.5ex] 
			9
		\end{array}
		\bigg )
		- 3 C_1 s \underline{u}
		\bigg ] .
		\label{eq:Xs0}
	\end{equation}
	Since $v_1$ only changes the constant $C_0$, we can choose $v_1 = 0$, leading to $\underline v = (0,9)$ [eq.~\eqref{eq:gev1}] and, via eq.~\eqref{eq:X0}, to
	\begin{equation}
		C_0 = Y_0, \qquad 
		C_1 = \frac{J_0 + 4 Y_0}{9} .
		\label{eq:C0C1}
	\end{equation} 
	When inserting eqs.~\eqref{eq:C0C1}, \eqref{eq:evu} into eq.~\eqref{eq:Xs0} we find the $s$ dependence (i.e.\ the time dependence) of the components $Y_s$ and $J_s$:
	\begin{equation}
		\begin{aligned}
			Y_s & = \left ( - \frac 13\right )^s \left  [ Y_0 - \frac s3 (J_0 + 4 Y_0 )\right] , \\[1ex]
			J_s & = \left ( - \frac 13\right )^s \left  [ J_0 + \frac{4 s}{3} (J_0 + 4 Y_0 )\right] ,
		\end{aligned}
		\label{eq:YsJs}
	\end{equation}
	where $Y_s = U_s / \tau$ with $\tau$ being the MD time-step. This gives back eq.~\ref{g25}
    upon using eq.~\ref{g24} as initial conditions.

\section{\label{AppA}How to find $\kappa^{*}$ and $\Lambda_{min}$}

We start with eq.~\ref{e33}.  It can be solved for the parameter $\kappa$ 
(considering the eigenvalue $\lambda$ as given):  
\begin{equation}\kappa =\left(1-\lambda\right)^3/\left(4\lambda^2-3\lambda +1\right
)\label{A1}\end{equation}
Setting $\lambda =1-1/x$ we get
\begin{equation}1/\kappa =f(x)\equiv 2x^3-5x^2+4x\label{A2}\end{equation}
For a given real $\kappa$ at least one root of eq.~\ref{e33} (considering $\lambda$ as 
unknown) must be real; let's denote it as 
\begin{equation}\lambda_0=1-1/x_0\label{A3}\end{equation}
Then $1/\kappa =f(x_0)$ with real $x_0$.  Eq.~\ref{A2} (for a given real $\kappa$) can 
be then simplified as 
\[2\left(x-x_0\right)\left(x^2+(x_0-5/2)x+f(x_0)/(2x_0)\right)=0\]
which leads to two further roots coming from the second factor in 
the above equation (in addition to $x=x_0$):
\begin{equation}x_{1,2}=\frac 12\left[\frac 52-x_0\pm\sqrt {3\left(7/6-x_0\right
)\left(x_0-1/2\right)}\right]\label{A4}\end{equation}
Equation~\ref{A4} shows that there is a repeated root $x_1 = x_2$, if $x_0 = 7/6$ or $x_0 = 1/2$. The root $x_0 = 1/2$ must be ruled out because eq.~\ref{A2} yields $\kappa = 1$, implying no stabilization of the iterative scheme. On the other hand, the root $x_0 = 7/6$ gives $x_1 = x_2 = 2/3$, leading to $\kappa = \kappa^* = 27/28$ [eq.~\ref{A2}] and to $\lambda_0 = 1/7$ and $\lambda_1 = \lambda_2 = - 1/2$ [eq.~\ref{A3}]. Hence,   the minimum of $\Lambda = \max(\left|\lambda_0\right|,\left|\lambda_
1\right|,\left|\lambda_2\right|) $ considered as a function of $\kappa $  is  $\Lambda_\text{min}= 1/2$ ; it is   achieved at $\kappa = \kappa^*$ in agreement with the discussion of Sec.~5.  

\begin{acknowledgments}
	We acknowledge funding by the French Agence Nationale de la Recherche (ANR) under Grant No.\ ANR-24-CE30-7668 (project VMFCMD).
\end{acknowledgments}

 %


\newpage

\Large \centerline{~Figure captions} \normalsize

Fig.  1:  Time dependencies of the reduced variables, 
$\mathcal A=AN\tau^2/2$, $\mathcal J=JN\tau$ and $\mathcal U=UN$, obtained 
using CMD based on the algorithm associated with eq.~\ref{g24} and 
the initial `configuration 4'.  Here $A$, $J$ and $U$ are the amplitudes of 
the external acceleration, the collective velocity and the collective 
displacement, respectively, at the selected wavevector $\underline q$, $\left|\underline 
q\right|=q_{min}$.  
The data are obtained for the 2D pLJ system with the total number 
of particles $N=10^4$ using MD time-step $\tau =0.005$.  All data are 
expressed in LJ units.  The theoretical results are shown by small 
squares, while the simulation data are indicated by solid lines. The 
upper graph shows real parts of transverse components of the reduced 
variables, while the lower graph exhibits the imaginary parts of
longitudinal components. 

Fig. 2:  Same as in figure 1, but for the initial `configuration 3'.

Fig.  3:  Same as in figure 1, but for a two-dimensional perfect gas whose initial 
microstate is identical to the `configuration 3' obtained for the pLJ 
system.  The second type of the CMD algorithm associated with 
eq.~\ref{g26} is used here; $\mathcal U^{wL}=U^{wL}N$ is the reduced 
longitudinal collective displacement whose initial value is defined by 
the mass density fluctuation (at the selected wavevector) implied by 
the initial configuration.  

Fig. 4:  Same as in figure 3, but for the 2D pLJ system instead of 
the perfect gas.

\begin{figure}
	\centering
	\setlength{\abovecaptionskip}{-20ex} 
	\includegraphics[height=1.1\textheight]{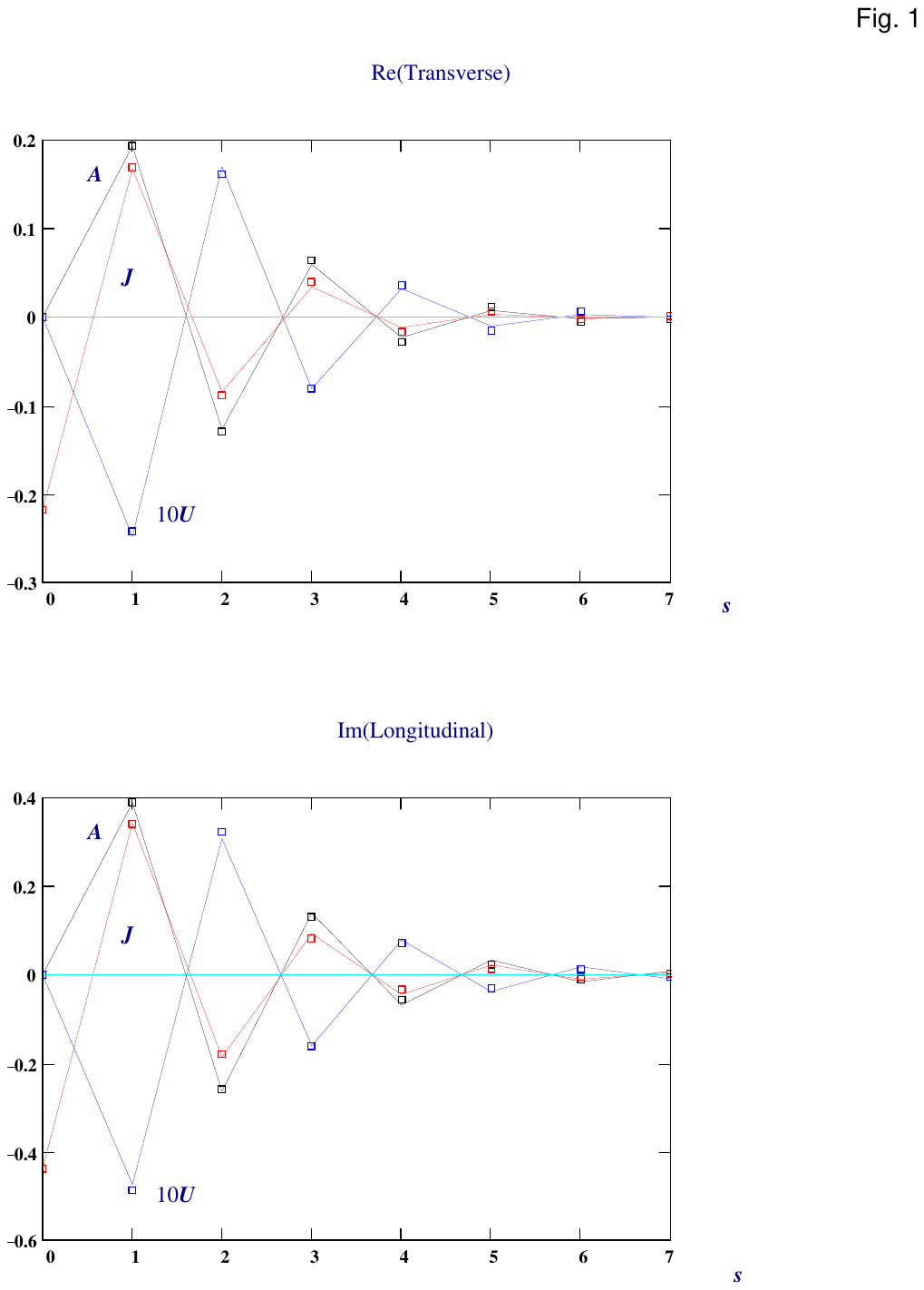}
	\caption{}
\end{figure}

\clearpage

\begin{figure}
	\centering
	\setlength{\abovecaptionskip}{-20ex} 
	\includegraphics[height=1.1\textheight]{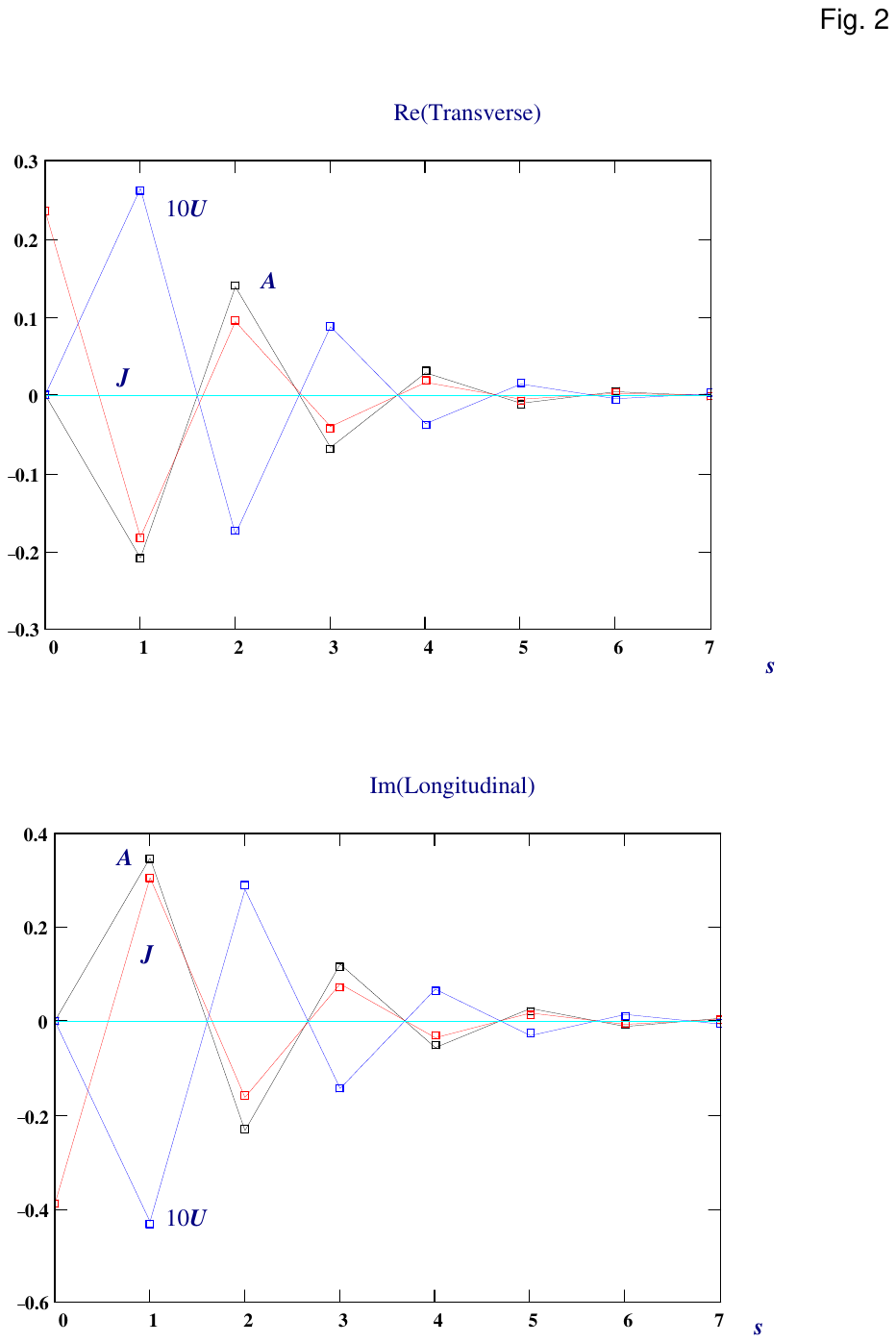}
	\caption{}
\end{figure}

\clearpage

\begin{figure}
	\centering
	\setlength{\abovecaptionskip}{-20ex} 
	\includegraphics[height=1.1\textheight]{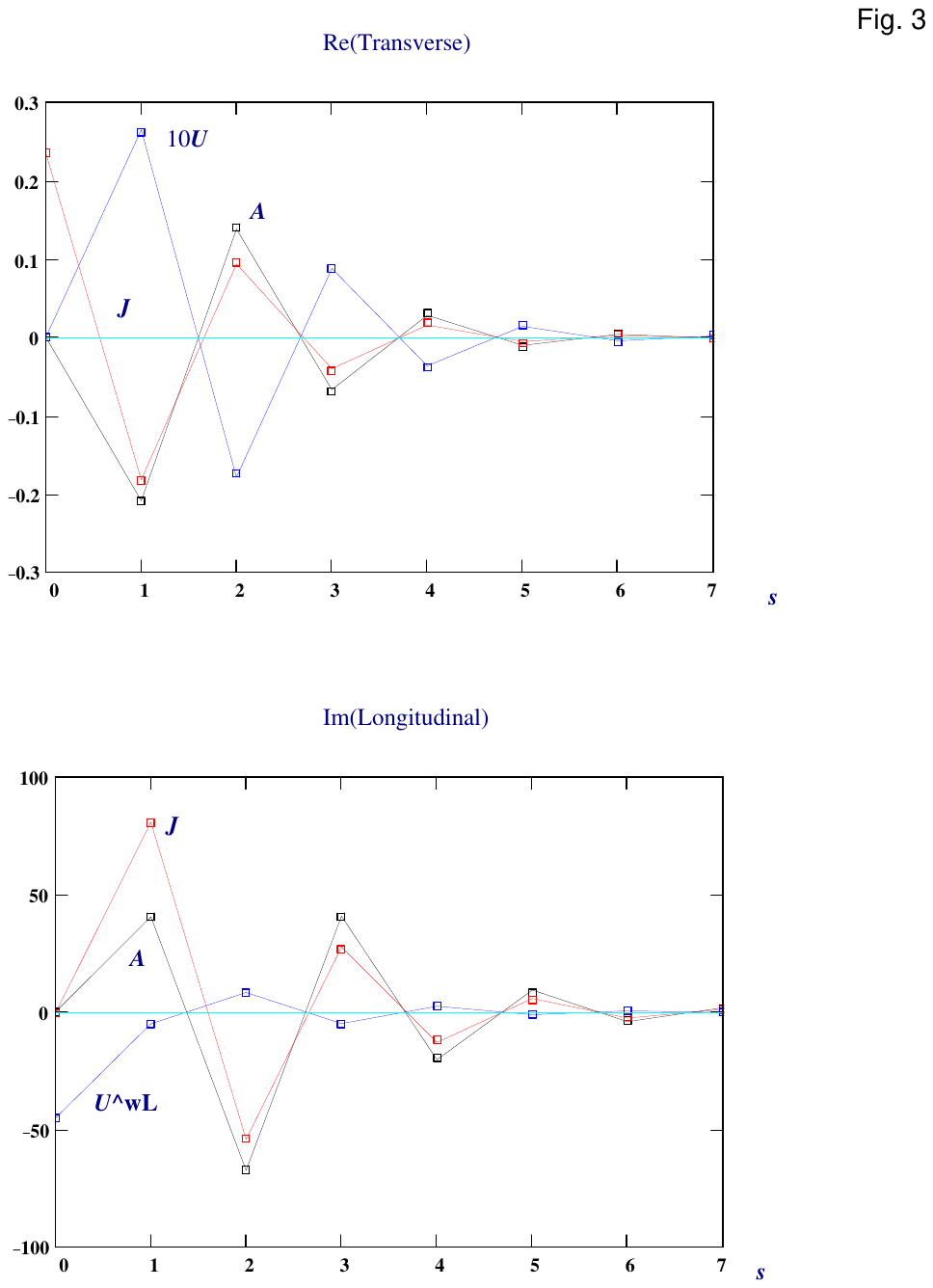}
	\caption{}
\end{figure}

\clearpage

\begin{figure}
	\centering
	\setlength{\abovecaptionskip}{-20ex} 
	\includegraphics[height=1.1\textheight]{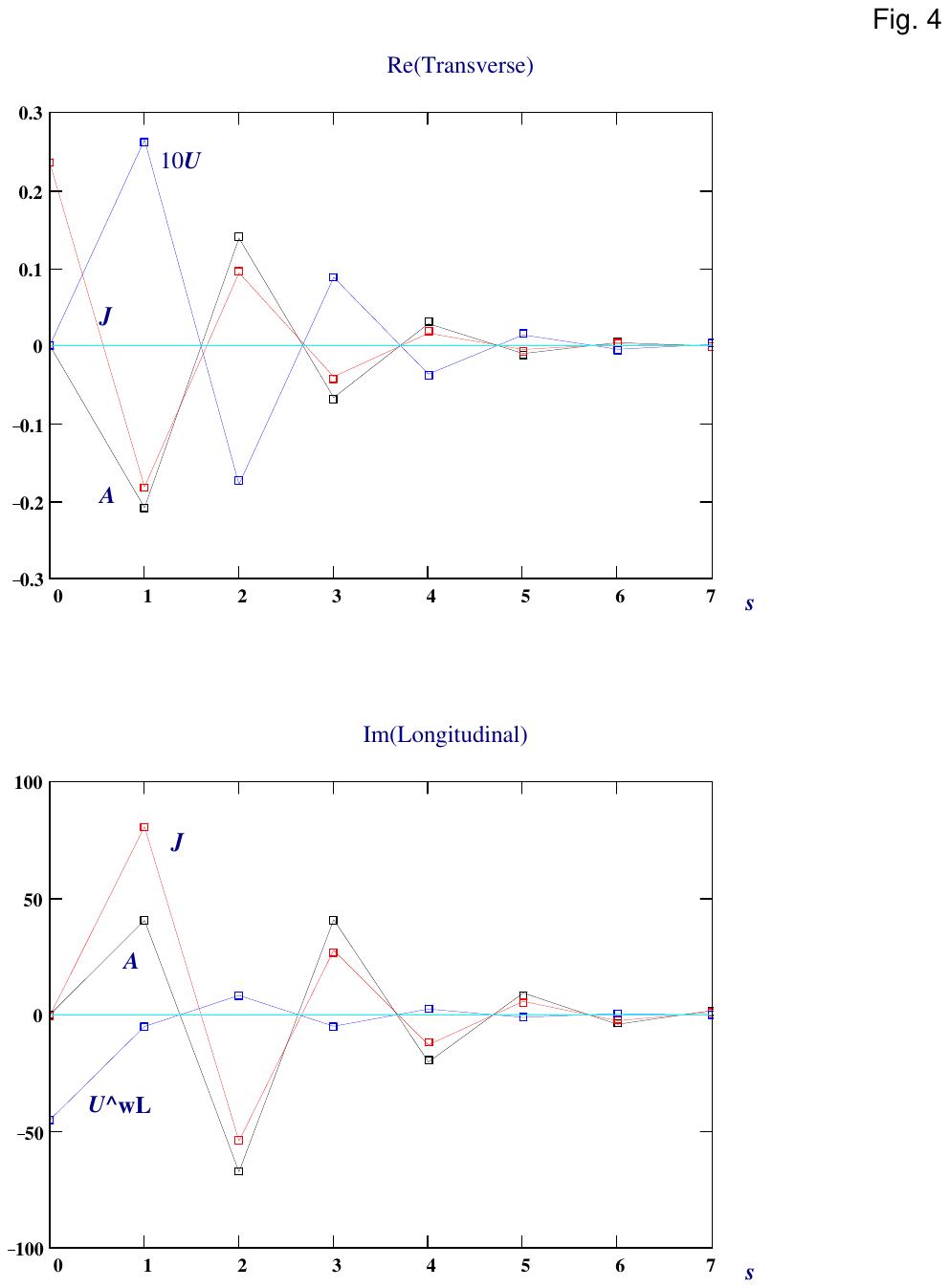}
	\caption{}
\end{figure}
 

\begin{thebibliography}{41}%

\bibitem{AlTil} M. P. Allen and D. J. Tildesley, Computer Simulation of
Liquids, Oxford University Press, Oxford, 2017.

\bibitem{xx1} I. Kriuchevskyi, J. P. Wittmer, H. Meyer and J. Baschnagel,
Phys. Rev. Lett., 2017, 119, 147802.

\bibitem{xx2} L. Klochko, J. Baschnagel, J. P. Wittmer and A. N. Semenov,
J. Chem. Phys., 2019, 151, 054504.

\bibitem{SM2021p7867} Klochko, L.; Baschnagel, J.;Wittmer, J.P.; Semenov, A.N.
Relaxation moduli of glass-forming systems: Temperature effects and
fluctuations. Soft Matter 2021, 17, 7867.
 
\bibitem{164505.17Deben} Debenedetti, P.G.; Stillinger, F.H.
Supercooled liquids and the glass transition. {\sl Nature}
  \mbox{2001}, {\sl 410}, 259--267.

\bibitem{N16} Gelin, S.; Tanaka, H.; Lema\^\i tre, A.  Anomalous 
phonon scattering and elastic correlations in amorphous solids.  {\sl Nat.
Mater.}  \mbox{2016}, \mbox{{\sl 15},  1177.}

\bibitem{N19} Hassani, M.; Zirdehi, E.M.; Kok, K.; Schall, P.; 
Fuchs, M.; Varnik, F.  Long-range strain correlations in 3d quiescent 
glass forming liquids.  {{\sl Europhys. Lett.}}
~\mbox{2018}, {\sl 124}, 18003.

\bibitem{N21} Illing, B.; Fritschi, S.; Hajnal, D.; Klix, C.; Keim, P.; Fuchs, M.  Strain pattern in supercooled liquids.  {\sl Phys.  Rev.
Lett.}          \mbox{2016}, \mbox{{\sl 117}, 208002.}

\bibitem{N6} Chacko, R.N.; Landes, F.P.; Biroli, G.; Dauchot, O.; Liu, A.; Reichman, D.R.  Elastoplasticity mediates dynamical 
heterogeneity below the mode coupling temperature. {\sl Phys.  Rev.
Lett.}          \mbox{2021}, {\sl 127}, 048002.

\bibitem{N24} Jung, G.; Biroli, G.; Berthier, L.  Predicting 
dynamic heterogeneity in glass-forming liquids by physics-inspired 
machine learning.  {\sl Phys.  Rev.  Lett.    \mbox{\/{\rm 2023}}}, {\sl 130}, 238202.

\bibitem{N5} Ozawa, M.; Biroli, G.  Elasticity, facilitation, and 
dynamic heterogeneity in glass-forming liquids.  {\sl Phys.  Rev.  Lett.}
        \mbox{2023}, {\sl 130}, 138201.

\bibitem{PRL2017} M. Maier, A. Zippelius and M. Fuchs, Phys. Rev. Lett., 2017,
119, 265701.


\bibitem{JCP2018} M. Maier, A. Zippelius and M. Fuchs, J. Chem. Phys., 2018,
149, 084502.

\bibitem{SM2018} L. Klochko, J. Baschnagel, J. P. Wittmer and A. N. Semenov,
Soft Matter, 2018, 14, 6835.

\bibitem{JCP2022} L.Klochko, J.Baschnagel, J.P.Wittmer, H. Meyer, O. Benzerara
and A. N. Semenov, J. Chem. Phys., 2022, 156, 164505.



\bibitem{Polymers2024} A. N. Semenov and J. Baschnagel, Polymers, 2024, 16, 2336.


\bibitem{SM2025} N.Grimm et al Soft Matter 21, 4256 (2025).

\bibitem{TadMil} E. B. Tadmor and R. E. Miller, Modeling Materials: Continuum,
Atomistic and Multiscale Techniques, Cambridge University
Press, Cambridge, 2014.


\bibitem{BalZop} U. Balucani and M. Zoppi, Dynamics of the Liquid State,
Oxford University Press, Oxford, 2003.


\bibitem{Phihl} I.Phihlajamaa et al. SciPost Phys. 15, 217 (2023).

\bibitem{WangZhan} Q.Wang, H.Zhan, Advances in Water Resources 75, 
80 (2015).

\bibitem{HansDon} J. P. Hansen and I. R. McDonald, Theory of Simple Liquids,
Academic Press, London, 1986.

\bibitem{kstress} O. Benzerara, J. Baschnagel, A.N. Semenov, L. Klochko, in preparation. 

 	\makeatletter
 	\providecommand \@ifxundefined [1]{%
 		\@ifx{#1\undefined}
 	}%
 	\providecommand \@ifnum [1]{%
 		\ifnum #1\expandafter \@firstoftwo
 		\else \expandafter \@secondoftwo
 		\fi
 	}%
 	\providecommand \@ifx [1]{%
 		\ifx #1\expandafter \@firstoftwo
 		\else \expandafter \@secondoftwo
 		\fi
 	}%
 	\providecommand \natexlab [1]{#1}%
 	\providecommand \enquote  [1]{``#1''}%
 	\providecommand \bibnamefont  [1]{#1}%
 	\providecommand \bibfnamefont [1]{#1}%
 	\providecommand \citenamefont [1]{#1}%
 	\providecommand \href@noop [0]{\@secondoftwo}%
 	\providecommand \href [0]{\begingroup \@sanitize@url \@href}%
 	\providecommand \@href[1]{\@@startlink{#1}\@@href}%
 	\providecommand \@@href[1]{\endgroup#1\@@endlink}%
 	\providecommand \@sanitize@url [0]{\catcode `\\12\catcode `\$12\catcode
 		`\&12\catcode `\#12\catcode `\^12\catcode `\_12\catcode `\%12\relax}%
 	\providecommand \@@startlink[1]{}%
 	\providecommand \@@endlink[0]{}%
 	\providecommand \url  [0]{\begingroup\@sanitize@url \@url }%
 	\providecommand \@url [1]{\endgroup\@href {#1}{\urlprefix }}%
 	\providecommand \urlprefix  [0]{URL }%
 	\providecommand \Eprint [0]{\href }%
 	\providecommand \doibase [0]{https://doi.org/}%
 	\providecommand \selectlanguage [0]{\@gobble}%
 	\providecommand \bibinfo  [0]{\@secondoftwo}%
 	\providecommand \bibfield  [0]{\@secondoftwo}%
 	\providecommand \translation [1]{[#1]}%
 	\providecommand \BibitemOpen [0]{}%
 	\providecommand \bibitemStop [0]{}%
 	\providecommand \bibitemNoStop [0]{.\EOS\space}%
 	\providecommand \EOS [0]{\spacefactor3000\relax}%
 	\providecommand \BibitemShut  [1]{\csname bibitem#1\endcsname}%
 	\let\auto@bib@innerbib\@empty
 	\bibitem [{\citenamefont {Berthier}\ and\ \citenamefont
 		{Biroli}(2011)}]{BerthierBiroliRMP}%
 	\BibitemOpen
 	\bibfield  {author} {\bibinfo {author} {\bibfnamefont {L.}~\bibnamefont
 			{Berthier}}\ and\ \bibinfo {author} {\bibfnamefont {G.}~\bibnamefont
 			{Biroli}},\ }\href@noop {} {\bibfield  {journal} {\bibinfo  {journal} {Rev.
 				Mod. Phys.}\ }\textbf {\bibinfo {volume} {83}},\ \bibinfo {pages} {587}
 		(\bibinfo {year} {2011})}\BibitemShut {NoStop}%
 	\bibitem [{\citenamefont {Fuchs}(2010)}]{Fuchs:AdvPolymSci2010}%
 	\BibitemOpen
 	\bibfield  {author} {\bibinfo {author} {\bibfnamefont {M.}~\bibnamefont
 			{Fuchs}},\ }\href@noop {} {\bibfield  {journal} {\bibinfo  {journal} {Adv.
 				Polym. Sci.}\ }\textbf {\bibinfo {volume} {36}},\ \bibinfo {pages} {55}
 		(\bibinfo {year} {2010})}\BibitemShut {NoStop}%
 	\bibitem [{\citenamefont {Nicolas}\ \emph {et~al.}(2018)\citenamefont
 		{Nicolas}, \citenamefont {Ferrero}, \citenamefont {Martens},\ and\
 		\citenamefont {Barrat}}]{NicolasEtal:RMP2018}%
 	\BibitemOpen
 	\bibfield  {author} {\bibinfo {author} {\bibfnamefont {A.}~\bibnamefont
 			{Nicolas}}, \bibinfo {author} {\bibfnamefont {E.~E.}\ \bibnamefont
 			{Ferrero}}, \bibinfo {author} {\bibfnamefont {K.}~\bibnamefont {Martens}},\
 		and\ \bibinfo {author} {\bibfnamefont {J.-L.}\ \bibnamefont {Barrat}},\
 	}\href@noop {} {\bibfield  {journal} {\bibinfo  {journal} {Rev. Mod. Phys.}\
 		}\textbf {\bibinfo {volume} {90}},\ \bibinfo {pages} {045006} (\bibinfo
 		{year} {2018})}\BibitemShut {NoStop}%
 	\bibitem [{\citenamefont {Voigtmann}(2014)}]{Voigtmann:COCIS2014}%
 	\BibitemOpen
 	\bibfield  {author} {\bibinfo {author} {\bibfnamefont {T.}~\bibnamefont
 			{Voigtmann}},\ }\href@noop {} {\bibfield  {journal} {\bibinfo  {journal}
 			{Curr. Opin. Colloid Interf. Sci.}\ }\textbf {\bibinfo {volume} {19}},\
 		\bibinfo {pages} {549} (\bibinfo {year} {2014})}\BibitemShut {NoStop}%
 	\bibitem [{\citenamefont {Allen}\ and\ \citenamefont
 		{Tildesley}(1987)}]{AllenTildesley}%
 	\BibitemOpen
 	\bibfield  {author} {\bibinfo {author} {\bibfnamefont {M.~P.}\ \bibnamefont
 			{Allen}}\ and\ \bibinfo {author} {\bibfnamefont {D.~J.}\ \bibnamefont
 			{Tildesley}},\ }\href@noop {} {\emph {\bibinfo {title} {Computer Simulation
 				of Liquids}}}\ (\bibinfo  {publisher} {Clarendon Press},\ \bibinfo {address}
 	{Oxford},\ \bibinfo {year} {1987})\BibitemShut {NoStop}%
 	\bibitem [{\citenamefont {Hansen}\ and\ \citenamefont
 		{McDonald}(1986)}]{HansenMcDonald}%
 	\BibitemOpen
 	\bibfield  {author} {\bibinfo {author} {\bibfnamefont {J.~P.}\ \bibnamefont
 			{Hansen}}\ and\ \bibinfo {author} {\bibfnamefont {I.~R.}\ \bibnamefont
 			{McDonald}},\ }\href@noop {} {\emph {\bibinfo {title} {Theory of Simple
 				Liquids}}}\ (\bibinfo  {publisher} {Academic Press},\ \bibinfo {address}
 	{London},\ \bibinfo {year} {1986})\BibitemShut {NoStop}%
 	\bibitem [{\citenamefont {G\"otze}(2009)}]{GoetzeBook2009}%
 	\BibitemOpen
 	\bibfield  {author} {\bibinfo {author} {\bibfnamefont {W.}~\bibnamefont
 			{G\"otze}},\ }\href@noop {} {\emph {\bibinfo {title} {Complex Dynamics of
 				Glass-Forming Liquids: A Mode-Coupling Theory}}}\ (\bibinfo  {publisher}
 	{Oxford University Press},\ \bibinfo {address} {Oxford},\ \bibinfo {year}
 	{2009})\BibitemShut {NoStop}%
 	\bibitem [{\citenamefont {Kriuchevskyi}\ \emph {et~al.}(2017)\citenamefont
 		{Kriuchevskyi}, \citenamefont {Wittmer}, \citenamefont {Meyer},\ and\
 		\citenamefont {Baschnagel}}]{KriuchevskyiEtal:PRL2017}%
 	\BibitemOpen
 	\bibfield  {author} {\bibinfo {author} {\bibfnamefont {I.}~\bibnamefont
 			{Kriuchevskyi}}, \bibinfo {author} {\bibfnamefont {J.~P.}\ \bibnamefont
 			{Wittmer}}, \bibinfo {author} {\bibfnamefont {H.}~\bibnamefont {Meyer}},\
 		and\ \bibinfo {author} {\bibfnamefont {J.}~\bibnamefont {Baschnagel}},\
 	}\href@noop {} {\bibfield  {journal} {\bibinfo  {journal} {Phys. Rev. Lett.}\
 		}\textbf {\bibinfo {volume} {119}},\ \bibinfo {pages} {147802} (\bibinfo
 		{year} {2017})}\BibitemShut {NoStop}%
 	\bibitem [{\citenamefont {Klochko}\ \emph {et~al.}(2021)\citenamefont
 		{Klochko}, \citenamefont {Baschnagel}, \citenamefont {Wittmer},\ and\
 		\citenamefont {Semenov}}]{KlochkoEtal:SoftMatter2021}%
 	\BibitemOpen
 	\bibfield  {author} {\bibinfo {author} {\bibfnamefont {L.}~\bibnamefont
 			{Klochko}}, \bibinfo {author} {\bibfnamefont {J.}~\bibnamefont {Baschnagel}},
 		\bibinfo {author} {\bibfnamefont {J.~P.}\ \bibnamefont {Wittmer}},\ and\
 		\bibinfo {author} {\bibfnamefont {A.~N.}\ \bibnamefont {Semenov}},\
 	}\href@noop {} {\bibfield  {journal} {\bibinfo  {journal} {Soft Matter}\
 		}\textbf {\bibinfo {volume} {17}},\ \bibinfo {pages} {7867} (\bibinfo {year}
 		{2021})}\BibitemShut {NoStop}%
 	\bibitem [{\citenamefont {Lema\^{\i}tre}(2014)}]{Lemaitre:PRL2014}%
 	\BibitemOpen
 	\bibfield  {author} {\bibinfo {author} {\bibfnamefont {A.}~\bibnamefont
 			{Lema\^{\i}tre}},\ }\href@noop {} {\bibfield  {journal} {\bibinfo  {journal}
 			{Phys. Rev. Lett.}\ }\textbf {\bibinfo {volume} {113}},\ \bibinfo {pages}
 		{245702} (\bibinfo {year} {2014})}\BibitemShut {NoStop}%
 	\bibitem [{\citenamefont {Lema\^{\i}tre}(2015)}]{Lemaitre:JCP2015}%
 	\BibitemOpen
 	\bibfield  {author} {\bibinfo {author} {\bibfnamefont {A.}~\bibnamefont
 			{Lema\^{\i}tre}},\ }\href@noop {} {\bibfield  {journal} {\bibinfo  {journal}
 			{J. Chem. Phys.}\ }\textbf {\bibinfo {volume} {143}},\ \bibinfo {pages}
 		{164515} (\bibinfo {year} {2015})}\BibitemShut {NoStop}%
 	\bibitem [{\citenamefont {Lema\^{\i}tre}(2017)}]{Lemaitre:PRE2017}%
 	\BibitemOpen
 	\bibfield  {author} {\bibinfo {author} {\bibfnamefont {A.}~\bibnamefont
 			{Lema\^{\i}tre}},\ }\href@noop {} {\bibfield  {journal} {\bibinfo  {journal}
 			{Phys. Rev. E}\ }\textbf {\bibinfo {volume} {96}},\ \bibinfo {pages} {052101}
 		(\bibinfo {year} {2017})}\BibitemShut {NoStop}%
 	\bibitem [{\citenamefont {Lema\^{\i}tre}(2018)}]{Lemaitre:JCP2018}%
 	\BibitemOpen
 	\bibfield  {author} {\bibinfo {author} {\bibfnamefont {A.}~\bibnamefont
 			{Lema\^{\i}tre}},\ }\href@noop {} {\bibfield  {journal} {\bibinfo  {journal}
 			{J. Chem. Phys.}\ }\textbf {\bibinfo {volume} {149}},\ \bibinfo {pages}
 		{104107} (\bibinfo {year} {2018})}\BibitemShut {NoStop}%
 	\bibitem [{\citenamefont {Wu}\ \emph {et~al.}(2015)\citenamefont {Wu},
 		\citenamefont {Iwashita},\ and\ \citenamefont {Egami}}]{WuEtal:PRE2015}%
 	\BibitemOpen
 	\bibfield  {author} {\bibinfo {author} {\bibfnamefont {B.}~\bibnamefont
 			{Wu}}, \bibinfo {author} {\bibfnamefont {T.}~\bibnamefont {Iwashita}},\ and\
 		\bibinfo {author} {\bibfnamefont {T.}~\bibnamefont {Egami}},\ }\href@noop {}
 	{\bibfield  {journal} {\bibinfo  {journal} {Phys. Rev. E}\ }\textbf {\bibinfo
 			{volume} {91}},\ \bibinfo {pages} {032301} (\bibinfo {year}
 		{2015})}\BibitemShut {NoStop}%
 	\bibitem [{\citenamefont {Chowdhury}\ \emph {et~al.}(2016)\citenamefont
 		{Chowdhury}, \citenamefont {Abraham}, \citenamefont {Hudson},\ and\
 		\citenamefont {Harrowell}}]{ChowdhuryEtal:JCP2016}%
 	\BibitemOpen
 	\bibfield  {author} {\bibinfo {author} {\bibfnamefont {S.}~\bibnamefont
 			{Chowdhury}}, \bibinfo {author} {\bibfnamefont {S.}~\bibnamefont {Abraham}},
 		\bibinfo {author} {\bibfnamefont {T.}~\bibnamefont {Hudson}},\ and\ \bibinfo
 		{author} {\bibfnamefont {P.}~\bibnamefont {Harrowell}},\ }\href@noop {}
 	{\bibfield  {journal} {\bibinfo  {journal} {J. Chem. Phys.}\ }\textbf
 		{\bibinfo {volume} {144}},\ \bibinfo {pages} {124508} (\bibinfo {year}
 		{2016})}\BibitemShut {NoStop}%
 	\bibitem [{\citenamefont {Tong}\ \emph {et~al.}(2020)\citenamefont {Tong},
 		\citenamefont {Sengupta},\ and\ \citenamefont
 		{Tanaka}}]{TongSenguptaTanaka:NatComm2020}%
 	\BibitemOpen
 	\bibfield  {author} {\bibinfo {author} {\bibfnamefont {H.}~\bibnamefont
 			{Tong}}, \bibinfo {author} {\bibfnamefont {S.}~\bibnamefont {Sengupta}},\
 		and\ \bibinfo {author} {\bibfnamefont {H.}~\bibnamefont {Tanaka}},\
 	}\href@noop {} {\bibfield  {journal} {\bibinfo  {journal} {Nat. Comm.}\
 		}\textbf {\bibinfo {volume} {11}},\ \bibinfo {pages} {4863} (\bibinfo {year}
 		{2020})}\BibitemShut {NoStop}%
 	\bibitem [{\citenamefont {Furukawa}\ and\ \citenamefont
 		{Tanaka}(2009)}]{FurukawaTanaka:PRL2009}%
 	\BibitemOpen
 	\bibfield  {author} {\bibinfo {author} {\bibfnamefont {A.}~\bibnamefont
 			{Furukawa}}\ and\ \bibinfo {author} {\bibfnamefont {H.}~\bibnamefont
 			{Tanaka}},\ }\href@noop {} {\bibfield  {journal} {\bibinfo  {journal} {Phys.
 				Rev. Lett.}\ }\textbf {\bibinfo {volume} {103}},\ \bibinfo {pages} {135703}
 		(\bibinfo {year} {2009})}\BibitemShut {NoStop}%
 	\bibitem [{\citenamefont {Maier}\ \emph {et~al.}(2018)\citenamefont {Maier},
 		\citenamefont {Zippelius},\ and\ \citenamefont {Fuchs}}]{MaierEtal:JCP2018}%
 	\BibitemOpen
 	\bibfield  {author} {\bibinfo {author} {\bibfnamefont {M.}~\bibnamefont
 			{Maier}}, \bibinfo {author} {\bibfnamefont {A.}~\bibnamefont {Zippelius}},\
 		and\ \bibinfo {author} {\bibfnamefont {M.}~\bibnamefont {Fuchs}},\
 	}\href@noop {} {\bibfield  {journal} {\bibinfo  {journal} {J. Chem. Phys.}\
 		}\textbf {\bibinfo {volume} {149}},\ \bibinfo {pages} {084502} (\bibinfo
 		{year} {2018})}\BibitemShut {NoStop}%
 	\bibitem [{\citenamefont {Vogel}\ and\ \citenamefont
 		{Fuchs}(2020)}]{VogelFuchs:EPJE2020}%
 	\BibitemOpen
 	\bibfield  {author} {\bibinfo {author} {\bibfnamefont {F.}~\bibnamefont
 			{Vogel}}\ and\ \bibinfo {author} {\bibfnamefont {M.}~\bibnamefont {Fuchs}},\
 	}\href@noop {} {\bibfield  {journal} {\bibinfo  {journal} {Eur. Phys. J. E}\
 		}\textbf {\bibinfo {volume} {43}},\ \bibinfo {pages} {70} (\bibinfo {year}
 		{2020})}\BibitemShut {NoStop}%
 	\bibitem [{\citenamefont {Steffen}\ \emph {et~al.}(2022)\citenamefont
 		{Steffen}, \citenamefont {Schneider}, \citenamefont {M\"uller},\ and\
 		\citenamefont {Rottler}}]{SteffenEtal:JCP2022}%
 	\BibitemOpen
 	\bibfield  {author} {\bibinfo {author} {\bibfnamefont {D.}~\bibnamefont
 			{Steffen}}, \bibinfo {author} {\bibfnamefont {L.}~\bibnamefont {Schneider}},
 		\bibinfo {author} {\bibfnamefont {M.}~\bibnamefont {M\"uller}},\ and\
 		\bibinfo {author} {\bibfnamefont {J.}~\bibnamefont {Rottler}},\ }\href@noop
 	{} {\bibfield  {journal} {\bibinfo  {journal} {J. Chem. Phys.}\ }\textbf
 		{\bibinfo {volume} {157}},\ \bibinfo {pages} {064501} (\bibinfo {year}
 		{2022})}\BibitemShut {NoStop}%
 	\bibitem [{\citenamefont {Klochko}\ \emph {et~al.}(2022)\citenamefont
 		{Klochko}, \citenamefont {Baschnagel}, \citenamefont {Wittmer}, \citenamefont
 		{Meyer}, \citenamefont {Benzerara},\ and\ \citenamefont
 		{Semenov}}]{KlochkoEtal:JCP2022}%
 	\BibitemOpen
 	\bibfield  {author} {\bibinfo {author} {\bibfnamefont {L.}~\bibnamefont
 			{Klochko}}, \bibinfo {author} {\bibfnamefont {J.}~\bibnamefont {Baschnagel}},
 		\bibinfo {author} {\bibfnamefont {J.~P.}\ \bibnamefont {Wittmer}}, \bibinfo
 		{author} {\bibfnamefont {H.}~\bibnamefont {Meyer}}, \bibinfo {author}
 		{\bibfnamefont {O.}~\bibnamefont {Benzerara}},\ and\ \bibinfo {author}
 		{\bibfnamefont {A.~N.}\ \bibnamefont {Semenov}},\ }\href@noop {} {\bibfield
 		{journal} {\bibinfo  {journal} {J. Chem. Phys.}\ }\textbf {\bibinfo {volume}
 			{156}},\ \bibinfo {pages} {164505} (\bibinfo {year} {2022})}\BibitemShut
 	{NoStop}%
 	\bibitem [{\citenamefont {Wittmer}\ \emph {et~al.}(2023)\citenamefont
 		{Wittmer}, \citenamefont {Semenov},\ and\ \citenamefont
 		{Baschnagel}}]{WittmerEtal_P5a:PRE2023}%
 	\BibitemOpen
 	\bibfield  {author} {\bibinfo {author} {\bibfnamefont {J.~P.}\ \bibnamefont
 			{Wittmer}}, \bibinfo {author} {\bibfnamefont {A.~N.}\ \bibnamefont
 			{Semenov}},\ and\ \bibinfo {author} {\bibfnamefont {J.}~\bibnamefont
 			{Baschnagel}},\ }\href@noop {} {\bibfield  {journal} {\bibinfo  {journal}
 			{Phys. Rev. E}\ }\textbf {\bibinfo {volume} {108}},\ \bibinfo {pages}
 		{015002} (\bibinfo {year} {2023})}\BibitemShut {NoStop}%
 	\bibitem [{\citenamefont {Grimm}\ \emph {et~al.}(2024)\citenamefont {Grimm},
 		\citenamefont {von Bischopinck}, \citenamefont {Zumbusch},\ and\
 		\citenamefont {Fuchs}}]{Grimm2024}%
 	\BibitemOpen
 	\bibfield  {author} {\bibinfo {author} {\bibfnamefont {N.}~\bibnamefont
 			{Grimm}}, \bibinfo {author} {\bibfnamefont {M.}~\bibnamefont {von
 				Bischopinck}}, \bibinfo {author} {\bibfnamefont {A.}~\bibnamefont
 			{Zumbusch}},\ and\ \bibinfo {author} {\bibfnamefont {M.}~\bibnamefont
 			{Fuchs}},\ }\href {https://doi.org/10.1063/5.0225890} {\bibfield  {journal}
 		{\bibinfo  {journal} {J. Chem. Phys.}\ }\textbf {\bibinfo {volume} {161}},\
 		\bibinfo {pages} {144118} (\bibinfo {year} {2024})}\BibitemShut {NoStop}%
 	\bibitem [{\citenamefont {Klochko}\ \emph {et~al.}(2018)\citenamefont
 		{Klochko}, \citenamefont {Baschnagel}, \citenamefont {Wittmer},\ and\
 		\citenamefont {Semenov}}]{KlochkoEtal:SoftMatter2018}%
 	\BibitemOpen
 	\bibfield  {author} {\bibinfo {author} {\bibfnamefont {L.}~\bibnamefont
 			{Klochko}}, \bibinfo {author} {\bibfnamefont {J.}~\bibnamefont {Baschnagel}},
 		\bibinfo {author} {\bibfnamefont {J.~P.}\ \bibnamefont {Wittmer}},\ and\
 		\bibinfo {author} {\bibfnamefont {A.~N.}\ \bibnamefont {Semenov}},\
 	}\href@noop {} {\bibfield  {journal} {\bibinfo  {journal} {Soft Matter}\
 		}\textbf {\bibinfo {volume} {14}},\ \bibinfo {pages} {6835} (\bibinfo {year}
 		{2018})}\BibitemShut {NoStop}%
 	\bibitem [{\citenamefont {Semenov}\ and\ \citenamefont
 		{Baschnagel}(2024)}]{SemenovBaschnagel:Polymers2024}%
 	\BibitemOpen
 	\bibfield  {author} {\bibinfo {author} {\bibfnamefont {A.~N.}\ \bibnamefont
 			{Semenov}}\ and\ \bibinfo {author} {\bibfnamefont {J.}~\bibnamefont
 			{Baschnagel}},\ }\href@noop {} {\bibfield  {journal} {\bibinfo  {journal}
 			{Polymers}\ }\textbf {\bibinfo {volume} {16}},\ \bibinfo {pages} {2336}
 		(\bibinfo {year} {2024})}\BibitemShut {NoStop}%
 	\bibitem [{\citenamefont {Martins}\ \emph {et~al.}(2023)\citenamefont
 		{Martins}, \citenamefont {Zhao}, \citenamefont {Demchuk}, \citenamefont
 		{Luo}, \citenamefont {Carden}, \citenamefont {Toleutay},\ and\ \citenamefont
 		{Sokolov}}]{MartinsEtal:Macro2023}%
 	\BibitemOpen
 	\bibfield  {author} {\bibinfo {author} {\bibfnamefont {M.~L.}\ \bibnamefont
 			{Martins}}, \bibinfo {author} {\bibfnamefont {X.}~\bibnamefont {Zhao}},
 		\bibinfo {author} {\bibfnamefont {Z.}~\bibnamefont {Demchuk}}, \bibinfo
 		{author} {\bibfnamefont {J.}~\bibnamefont {Luo}}, \bibinfo {author}
 		{\bibfnamefont {G.~P.}\ \bibnamefont {Carden}}, \bibinfo {author}
 		{\bibfnamefont {G.}~\bibnamefont {Toleutay}},\ and\ \bibinfo {author}
 		{\bibfnamefont {A.~P.}\ \bibnamefont {Sokolov}},\ }\href@noop {} {\bibfield
 		{journal} {\bibinfo  {journal} {Macromolecules}\ }\textbf {\bibinfo {volume}
 			{56}},\ \bibinfo {pages} {8688} (\bibinfo {year} {2023})}\BibitemShut
 	{NoStop}%
 	\bibitem [{\citenamefont {Likhtman}(2012)}]{Likhtman:Review2012}%
 	\BibitemOpen
 	\bibfield  {author} {\bibinfo {author} {\bibfnamefont {A.~E.}\ \bibnamefont
 			{Likhtman}},\ }\bibinfo {title} {Polymer science: A comprehensive reference}\
 	(\bibinfo  {publisher} {Elsevier},\ \bibinfo {address} {Amsterdam},\ \bibinfo
 	{year} {2012})\ Chap.\ \bibinfo {chapter} {Viscoelasticity and Molecular
 		Rheology}, pp.\ \bibinfo {pages} {133--179}\BibitemShut {NoStop}%
 	\bibitem [{\citenamefont {Semenov}\ \emph {et~al.}(2012)\citenamefont
 		{Semenov}, \citenamefont {Farago},\ and\ \citenamefont
 		{Meyer}}]{SemenovEtal:JCP2012}%
 	\BibitemOpen
 	\bibfield  {author} {\bibinfo {author} {\bibfnamefont {A.~N.}\ \bibnamefont
 			{Semenov}}, \bibinfo {author} {\bibfnamefont {J.}~\bibnamefont {Farago}},\
 		and\ \bibinfo {author} {\bibfnamefont {H.}~\bibnamefont {Meyer}},\
 	}\href@noop {} {\bibfield  {journal} {\bibinfo  {journal} {J. Chem. Phys.}\
 		}\textbf {\bibinfo {volume} {136}},\ \bibinfo {pages} {244905} (\bibinfo
 		{year} {2012})}\BibitemShut {NoStop}%
 	\bibitem [{\citenamefont {Chacko}\ \emph {et~al.}(2021)\citenamefont {Chacko},
 		\citenamefont {Landes}, \citenamefont {Biroli}, \citenamefont {Dauchot},
 		\citenamefont {Liu},\ and\ \citenamefont {Reichman}}]{ChackoEtal:PRL2021}%
 	\BibitemOpen
 	\bibfield  {author} {\bibinfo {author} {\bibfnamefont {R.~N.}\ \bibnamefont
 			{Chacko}}, \bibinfo {author} {\bibfnamefont {F.~P.}\ \bibnamefont {Landes}},
 		\bibinfo {author} {\bibfnamefont {G.}~\bibnamefont {Biroli}}, \bibinfo
 		{author} {\bibfnamefont {O.}~\bibnamefont {Dauchot}}, \bibinfo {author}
 		{\bibfnamefont {A.~J.}\ \bibnamefont {Liu}},\ and\ \bibinfo {author}
 		{\bibfnamefont {D.~R.}\ \bibnamefont {Reichman}},\ }\href@noop {} {\bibfield
 		{journal} {\bibinfo  {journal} {Phys. Rev. Lett.}\ }\textbf {\bibinfo
 			{volume} {127}},\ \bibinfo {pages} {048002} (\bibinfo {year}
 		{2021})}\BibitemShut {NoStop}%
 	\bibitem [{\citenamefont {Ozawa}\ and\ \citenamefont
 		{Biroli}(2023)}]{OzawaBiroli:PRL2023}%
 	\BibitemOpen
 	\bibfield  {author} {\bibinfo {author} {\bibfnamefont {M.}~\bibnamefont
 			{Ozawa}}\ and\ \bibinfo {author} {\bibfnamefont {G.}~\bibnamefont {Biroli}},\
 	}\href@noop {} {\bibfield  {journal} {\bibinfo  {journal} {Phys. Rev. Lett.}\
 		}\textbf {\bibinfo {volume} {130}},\ \bibinfo {pages} {138201} (\bibinfo
 		{year} {2023})}\BibitemShut {NoStop}%
 	\bibitem [{\citenamefont {Maier}\ \emph {et~al.}(2017)\citenamefont {Maier},
 		\citenamefont {Zippelius},\ and\ \citenamefont {Fuchs}}]{MaierEtal:PRL2017}%
 	\BibitemOpen
 	\bibfield  {author} {\bibinfo {author} {\bibfnamefont {M.}~\bibnamefont
 			{Maier}}, \bibinfo {author} {\bibfnamefont {A.}~\bibnamefont {Zippelius}},\
 		and\ \bibinfo {author} {\bibfnamefont {M.}~\bibnamefont {Fuchs}},\
 	}\href@noop {} {\bibfield  {journal} {\bibinfo  {journal} {Phys. Rev. Lett.}\
 		}\textbf {\bibinfo {volume} {119}},\ \bibinfo {pages} {265701} (\bibinfo
 		{year} {2017})}\BibitemShut {NoStop}%
 	\bibitem [{\citenamefont {Maier}\ \emph {et~al.}(2024)\citenamefont {Maier},
 		\citenamefont {Zippelius},\ and\ \citenamefont {Fuchs}}]{MaierEtal:JCP2024}%
 	\BibitemOpen
 	\bibfield  {author} {\bibinfo {author} {\bibfnamefont {M.}~\bibnamefont
 			{Maier}}, \bibinfo {author} {\bibfnamefont {A.}~\bibnamefont {Zippelius}},\
 		and\ \bibinfo {author} {\bibfnamefont {M.}~\bibnamefont {Fuchs}},\
 	}\href@noop {} {\bibfield  {journal} {\bibinfo  {journal} {J. Chem. Phys.}\
 		}\textbf {\bibinfo {volume} {161}},\ \bibinfo {pages} {199901} (\bibinfo
 		{year} {2024})}\BibitemShut {NoStop}%
 	\bibitem [{\citenamefont {Baschnagel}\ and\ \citenamefont
 		{Semenov}(2026)}]{BaschnagelSemenov:SoftMatter2026}%
 	\BibitemOpen
 	\bibfield  {author} {\bibinfo {author} {\bibfnamefont {J.}~\bibnamefont
 			{Baschnagel}}\ and\ \bibinfo {author} {\bibfnamefont {A.~N.}\ \bibnamefont
 			{Semenov}},\ }\href@noop {} {\bibfield  {journal} {\bibinfo  {journal} {Soft
 				Matter}\ }\textbf {\bibinfo {volume} {22}},\ \bibinfo {pages} {4832}
 		(\bibinfo {year} {2026})}\BibitemShut {NoStop}%
 	\bibitem [{\citenamefont {Janssen}(2018)}]{Janssen2018}%
 	\BibitemOpen
 	\bibfield  {author} {\bibinfo {author} {\bibfnamefont {L.~M.~C.}\
 			\bibnamefont {Janssen}},\ }\href@noop {} {\bibfield  {journal} {\bibinfo
 			{journal} {Front. Phys.}\ }\textbf {\bibinfo {volume} {6}},\ \bibinfo {pages}
 		{97} (\bibinfo {year} {2018})}\BibitemShut {NoStop}%
 	\bibitem [{\citenamefont {Pihlajamaa}\ \emph {et~al.}(2023)\citenamefont
 		{Pihlajamaa}, \citenamefont {Debets}, \citenamefont {Laudicina},\ and\
 		\citenamefont {Janssen}}]{PihlajamaaEtal:preprint2023}%
 	\BibitemOpen
 	\bibfield  {author} {\bibinfo {author} {\bibfnamefont {I.}~\bibnamefont
 			{Pihlajamaa}}, \bibinfo {author} {\bibfnamefont {V.~E.}\ \bibnamefont
 			{Debets}}, \bibinfo {author} {\bibfnamefont {C.~C.~L.}\ \bibnamefont
 			{Laudicina}},\ and\ \bibinfo {author} {\bibfnamefont {L.~M.~C.}\ \bibnamefont
 			{Janssen}},\ }\href@noop {} {\bibfield  {journal} {\bibinfo  {journal}
 			{SciPost Phys.}\ }\textbf {\bibinfo {volume} {15}},\ \bibinfo {pages} {217}
 		(\bibinfo {year} {2023})}\BibitemShut {NoStop}%
 	\bibitem [{\citenamefont {Pihlajamaa}\ and\ \citenamefont
 		{Janssen}(2024)}]{PihlajamaaJanssen:PRR2024}%
 	\BibitemOpen
 	\bibfield  {author} {\bibinfo {author} {\bibfnamefont {I.}~\bibnamefont
 			{Pihlajamaa}}\ and\ \bibinfo {author} {\bibfnamefont {L.~M.~C.}\ \bibnamefont
 			{Janssen}},\ }\href@noop {} {\bibfield  {journal} {\bibinfo  {journal} {Phys.
 				Rev. Res.}\ }\textbf {\bibinfo {volume} {6}},\ \bibinfo {pages} {043319}
 		(\bibinfo {year} {2024})}\BibitemShut {NoStop}%
 	\bibitem [{\citenamefont {Baity-Jesi}\ and\ \citenamefont
 		{Reichman}(2019)}]{Baity-JesiReichman:JCP2019}%
 	\BibitemOpen
 	\bibfield  {author} {\bibinfo {author} {\bibfnamefont {M.}~\bibnamefont
 			{Baity-Jesi}}\ and\ \bibinfo {author} {\bibfnamefont {D.~R.}\ \bibnamefont
 			{Reichman}},\ }\href@noop {} {\bibfield  {journal} {\bibinfo  {journal} {J.
 				Chem. Phys.}\ }\textbf {\bibinfo {volume} {151}},\ \bibinfo {pages} {084503}
 		(\bibinfo {year} {2019})}\BibitemShut {NoStop}%
 	\bibitem [{\citenamefont {Grimm}\ \emph {et~al.}(2025)\citenamefont {Grimm},
 		\citenamefont {Baschnagel}, \citenamefont {Semenov}, \citenamefont
 		{Zippelius},\ and\ \citenamefont {Fuchs}}]{GrimmEtal:SoftMatter2025}%
 	\BibitemOpen
 	\bibfield  {author} {\bibinfo {author} {\bibfnamefont {N.}~\bibnamefont
 			{Grimm}}, \bibinfo {author} {\bibfnamefont {J.}~\bibnamefont {Baschnagel}},
 		\bibinfo {author} {\bibfnamefont {A.~N.}\ \bibnamefont {Semenov}}, \bibinfo
 		{author} {\bibfnamefont {A.}~\bibnamefont {Zippelius}},\ and\ \bibinfo
 		{author} {\bibfnamefont {M.}~\bibnamefont {Fuchs}},\ }\href@noop {}
 	{\bibfield  {journal} {\bibinfo  {journal} {Soft Matter}\ }\textbf {\bibinfo
 			{volume} {21}},\ \bibinfo {pages} {4256} (\bibinfo {year}
 		{2025})}\BibitemShut {NoStop}%
 	\bibitem [{\citenamefont {Benzerara}\ \emph {et~al.}(view)\citenamefont
 		{Benzerara}, \citenamefont {Baschnagel},\ and\ \citenamefont
 		{Semenov}}]{BenzeraraSemenovBaschnagel:PRL2026}%
 	\BibitemOpen
 	\bibfield  {author} {\bibinfo {author} {\bibfnamefont {O.}~\bibnamefont
 			{Benzerara}}, \bibinfo {author} {\bibfnamefont {J.}~\bibnamefont
 			{Baschnagel}},\ and\ \bibinfo {author} {\bibfnamefont {A.~N.}\ \bibnamefont
 			{Semenov}},\ }\href@noop {} {\bibfield  {journal} {\bibinfo  {journal} {XXX}\
 		}\textbf {\bibinfo {volume} {XX}},\ \bibinfo {pages} {XXXX} (\bibinfo {year}
 		{under review})}\BibitemShut {NoStop}%
 	\bibitem [{\citenamefont {Tanguy}\ \emph {et~al.}(2002)\citenamefont {Tanguy},
 		\citenamefont {Wittmer}, \citenamefont {L\'eonforte},\ and\ \citenamefont
 		{Barrat}}]{TanguyEtal:PRB2002}%
 	\BibitemOpen
 	\bibfield  {author} {\bibinfo {author} {\bibfnamefont {A.}~\bibnamefont
 			{Tanguy}}, \bibinfo {author} {\bibfnamefont {J.~P.}\ \bibnamefont {Wittmer}},
 		\bibinfo {author} {\bibfnamefont {F.}~\bibnamefont {L\'eonforte}},\ and\
 		\bibinfo {author} {\bibfnamefont {J.-L.}\ \bibnamefont {Barrat}},\
 	}\href@noop {} {\bibfield  {journal} {\bibinfo  {journal} {Phys. Rev. B}\
 		}\textbf {\bibinfo {volume} {66}},\ \bibinfo {pages} {174205} (\bibinfo
 		{year} {2002})}\BibitemShut {NoStop}%
 	\bibitem [{\citenamefont {Wittmer}\ \emph {et~al.}(2013)\citenamefont
 		{Wittmer}, \citenamefont {Xu}, \citenamefont {Poli{\'n}ska}, \citenamefont
 		{Weysser},\ and\ \citenamefont {Baschnagel}}]{WittmerEtal:JCP2013}%
 	\BibitemOpen
 	\bibfield  {author} {\bibinfo {author} {\bibfnamefont {J.~P.}\ \bibnamefont
 			{Wittmer}}, \bibinfo {author} {\bibfnamefont {H.}~\bibnamefont {Xu}},
 		\bibinfo {author} {\bibfnamefont {P.}~\bibnamefont {Poli{\'n}ska}}, \bibinfo
 		{author} {\bibfnamefont {F.}~\bibnamefont {Weysser}},\ and\ \bibinfo {author}
 		{\bibfnamefont {J.}~\bibnamefont {Baschnagel}},\ }\href@noop {} {\bibfield
 		{journal} {\bibinfo  {journal} {J. Chem.\ Phys.}\ }\textbf {\bibinfo {volume}
 			{138}},\ \bibinfo {pages} {12A533} (\bibinfo {year} {2013})}\BibitemShut
 	{NoStop}%
 \end{thebibliography}
\end{document}